\documentclass[10pt]{article}
\usepackage{framed}
\usepackage{empheq,nccmath}
\usepackage{dsfont}
\usepackage{amsmath}
\usepackage{amssymb}
\usepackage{calrsfs}
\usepackage{graphicx}
\usepackage{subcaption}
\usepackage{enumitem}
\usepackage{mathrsfs}
\usepackage{cite}
\usepackage[left=2.1cm,right=2.1cm,top=1.2cm,bottom=1.3cm,includeheadfoot]{geometry}
\usepackage{indentfirst}
\usepackage[dvipsnames]{xcolor}
\usepackage[colorlinks=true,linkcolor=black,citecolor=black]{hyperref}
\usepackage[notlot,notlof,nottoc]{tocbibind}

\begin{document}
\begin{center}
	{\Large\bf
	{Hybrid classical-quantum systems in terms of moments}\\
	}
	
	\vskip 4mm
	
	David Brizuela\footnote{ david.brizuela@ehu.eus}
	and
	Sara F. Uria\footnote{sara.fernandezu@ehu.eus}
	
	\vskip 3mm
	{\sl Department of Physics and EHU Quantum Center, University of the Basque Country UPV/EHU,\\
		Barrio Sarriena s/n, 48940 Leioa, Spain}
	
\end{center}
\begin{abstract}
We present a consistent formalism to describe the dynamics of hybrid systems with
mixed classical and quantum degrees of freedom.
The probability function of the system, which, in general,
will be a combination of the classical distribution function and the quantum density matrix,
is described in terms of its corresponding moments.
We then define a hybrid Poisson bracket, such that the dynamics of the moments is
ruled by an effective Hamiltonian. In particular, a closed formula
for the Poisson brackets between any two moments for an arbitrary number of degrees of freedom
is presented, which corrects previous expressions derived in the literature for the purely quantum case.
This formula is of special relevance for practical applications of the formalism.
Finally, we study the dynamics of a particular hybrid system given by two coupled oscillators,
one being quantum and the other classical.
Due to the coupling, specific quantum and classical properties are transferred between different sectors.
In particular, the quantum sector is allowed to violate the uncertainty relation, though we explicitly show that
there exists a minimum positive bound of the total uncertainty of the hybrid system.

\end{abstract}

\section{Introduction}\label{sec:intro}
The boundary between classical and quantum physics remains unknown. It is within this context that the exploration of hybrid classical-quantum physics has emerged, aiming to describe the interaction between genuinely classical and
genuinely quantum systems.  
Certainly, this study can shed light on the aforementioned problem and, moreover, it is motivated by a variety of reasons, both practical and theoretical. For instance, in the Copenhagen interpretation of quantum
mechanics, measuring devices are regarded as classical \cite{cop1,cop2,cop3,cop4}. Therefore, a comprehensive understanding of the coexistence of quantum and classical sectors is essential for this foundational interpretation. Furthermore, if one considers that gravity is fundamentally classical, hybrid classical-quantum physics can be essential to understand how it interacts with quantum systems \cite{grav,Oppenheim:2018igd}. On the other hand,
from a practical standpoint, in certain intricate quantum systems, considering some degrees of freedom as classical can be very useful in order to obtain a simpler approximate description of the dynamics. This is of particular relevance in fields like molecular or condensed matter physics \cite{cond1,cond2,cond3,cond4,cond5,cond6,cond7}, but it is also the main motivation to consider
	quantum field theory on classical curved backgrounds as a good approximation to a prospective full theory of quantum gravity.

Undoubtedly, these examples highlight the necessity for a hybrid theory. However, so far there is not a unique, perfect, construction, and several proposals can be found in the literature (for reviews
on the different proposals we refer the reader to Refs. \cite{Terno:2023hvq, BCG12}). These approaches exhibit a broad spectrum of methodologies.  For example, some try to preserve the use of quantum states and trajectories to describe, respectively, the quantum and classical sectors \cite{grav,cat1}. Moreover, some focus on a quantum formalism for the whole hybrid system, by formulating the classical sector as quantum \cite{cat2_1,cat2_2,cat2_3,cat2_4,cat2_5,cat2_6,Peres_Terno,cat2_7,Oppenheim:2018igd,Aleksandrov,Gerasimenko,Blanchard_Jadczyk,Diosi,Gay-Balmaz:2021zve,BermudezManjarres:2021ago,Bondar_2019}. Conversely, some others formulate the quantum sector in classical terms, in order to work with a formally complete classical system \cite{cat3_1,cat3_2,cat3_3,cat3_4,cat3_5,cat3_6}. Additionally, certain proposals bring the quantum and classical sectors into a common language, in order to extend it to the interacting hybrid system \cite{cat4_1,cat4_2,cat4_3,cat4_4}. It is also worth mentioning that some approaches begin by building a fully quantum system and then take a classical limit in one of the subsystems, in order to get a classical-quantum interacting system \cite{Oliynyk:2015ngj}.
Nevertheless, all of them deal with some inconsistencies, such as the inability to define a Lie bracket, the nonpositivity of the probability distribution, an inconsistent classical limit,
a nonunitary evolution, or the inability to distinguish between the classical and quantum sectors. 

Hence, in this paper, we present a new proposal for a consistent description of the dynamics of hybrid systems, with a different starting point and formalism. It is motivated by a formalism for quantum systems that replaces the wave function (or more generally the density matrix), by its infinite set of moments \cite{BM98, BS06}.
As analyzed in Refs. \cite{BM98, Brizuela:2014cfa}, this approach can also be used to describe
classical ensembles, and, thus, we propose a similar extension of the formalism to include hybrid systems. Such extension will rely on the definition of a bracket for hybrid observables, which will
generalize the quantum commutator and the classical Poisson bracket.

The rest of the paper is organized as follows. In Sec. \ref{sec:quantum}
the formalism for quantum systems is presented. As commented above, this is a well known formalism, though, as an original result, here we provide a closed formula
for the Poisson bracket between moments for any number of degrees of freedom, which corrects the formula presented in Ref. \cite{BS06}.
In Sec. \ref{sec:classical} we show that a similar formalism as presented
for quantum systems can be used to describe the dynamics of classical ensembles.
This generalizes the work of Refs. \cite{BM98, Brizuela:2014cfa}
to an arbitrary number of degrees of freedom. In Sec. \ref{sec:hybrid}
we present the hybrid case, with mixed classical and quantum degrees of freedom,
and, in particular, define the hybrid bracket that generalizes the quantum commutator
and the classical Poisson bracket. Sec. \ref{sec:harmonic} comments on the specific
dynamical properties of harmonic Hamiltonians, defined as those that are at
most quadratic on basic variables. Then, in Sec. \ref{sec:application} a particular
application of the formalism is considered, and we study in detail the hybrid
system given by a classical oscillator coupled to a quantum one. Finally,
in Sec. \ref{sec:conclusions} we present the conclusions and summarize the main results of the paper.
There are two additional appendices with technical computations: in App.~\ref{app:Poisson}
we derive the main steps of the derivation of the formula for the Poisson bracket between moments, while in
App.~\ref{app:Lie} the properties of the bracket between hybrid observables are analyzed.

\section{Quantum Systems}\label{sec:quantum}

Let us consider a system with $N$ quantum degrees of freedom described by
the basic operators $(\hat{q}_j,\hat{p}_j)_{j=1}^N$, which obey the usual canonical commutation relations $[\hat{q}_j,\hat{p}_k]=i\hbar\delta_{jk}$.
The central moments of the wave function, or, more generically, of the density matrix, are defined as the expectation value
\begin{align}\label{def_moments}
\Delta\big(q_1^{a_1}p_1^{b_1}\dots q_N^{a_N}p_N^{b_N}\big):=
\left\langle
(\hat{q}_1-q_1)^{a_1}
(\hat{p}_1-p_1)^{b_1}
\dots
(\hat{q}_N-q_N)^{a_N}
(\hat{p}_N-p_N)^{b_N}
\right\rangle_{\text{Weyl}},
\end{align}
for any nonnegative integers $a_i$, $b_i$. In this expression we have defined $q_i:=\langle \hat{q}_i\rangle$ and
$p_i:=\langle \hat{p}_i\rangle$, while the subscript Weyl refers to a totally symmetric ordering of the basic operators.
A priori one could choose another ordering for this definition but, with the present choice, one ensures
that all the moments are real-valued time-dependent functions. Therefore, this definition can be understood as a decomposition
of the density matrix of the system into an infinite set of variables, which only
depend on time and represent the actual measurable quantities of the system.
We note that, in the above definition, we are implicitly assuming that
the corresponding expectation values are well defined for the state under consideration.

In the Heisenberg picture the state of the system is described by a constant (time-independent)
wave function (or density matrix), and the evolution of any operator $\hat A$ follows
\begin{align}\label{heisenberg_evolution}
\frac{d \hat A}{dt}=\frac{\partial \hat A}{\partial t}+\frac{1}{i\hbar}\left[\hat A,\hat H\right],
\end{align}
where $\hat H$ is the Hamiltonian operator. Taking the expectation value of this expression, one obtains
\begin{align}\label{ehrenfest_1}
\frac{d\langle\hat A\rangle}{dt}=\bigg\langle\frac{\partial \hat A}{\partial t}\bigg\rangle+
\frac{1}{i\hbar}\left\langle\left[\hat A,\hat H\right]\right\rangle.
\end{align}
Then, defining the Poisson bracket between expectation values as $\{\langle\hat{f}\rangle,\langle\hat{g}\rangle\}:=-i\hbar^{-1}\langle
[\hat{f},\hat{g}]\rangle$ leads to
\begin{align}\label{ehrenfest_2}
\frac{d\langle\hat A\rangle}{dt}=\bigg\langle\frac{\partial \hat A}{\partial t}\bigg\rangle+\big\{\langle\hat A\rangle,\langle\hat H\rangle\big\}.
\end{align}
Since the total time derivative of $\langle \hat A \rangle$
is thus given by its partial time derivative plus its Poisson bracket with $\langle \hat H\rangle$,
this object can be identified as the Hamiltonian that rules the dynamics
of the expectation value of any operator $\hat A$. Therefore, we define $H_Q:=\langle \hat H\rangle$.
If we then formally expand this Hamiltonian around the expectation values $(q_i,p_i)$,
it is possible to write it in terms of the moments as follows,
\begin{align}
H_Q:=&\big\langle{\hat{H}(t,\hat q_1,\hat p_1,\dots,\hat q_N,\hat p_N)}\big\rangle=
 \big\langle{\hat{H}(t,\hat q_1-q_1+q_1,\hat p_1-p_1+p_1,\dots,\hat q_N-q_N+q_N,\hat p_N-p_N+p_N)}\big\rangle
\nonumber
\\
=&H(t,q_1,p_1,\dots,q_N,p_N)+\sum_{J=2}^{+\infty}
\left(\prod_{k=1}^{N}\frac{1}{m_k!}\frac{1}{n_k!}\right)
\dfrac{\partial^{J}H}
{\partial q_1^{m_1}\partial p_1^{n_1}\dots\partial q_N^{m_N}p_N^{n_N}}
\Delta(q_1^{m_1}p_1^{n_1}\dots q_N^{m_N}p_N^{n_N}),
\label{effective Hamiltonian}
\end{align}
where $J=m_1+n_1+\dots+m_N+n_N$, and, for each value of $J$, all possible nonnegative integers $m_j$ must be considered.
Moreover, $H(t,q_1,p_1,\dots,q_N,p_N)$ is the classical Hamiltonian,
which is obtained just by replacing the operators $(\hat{q}_i,\hat{p}_i)$ by their corresponding expectation values
$(q_i,p_i)$ in the explicit expression of the Hamiltonian operator $\hat{H}=\hat{H}(t,\hat q_1,\hat p_1,\dots,\hat q_N,\hat p_N)$.
In order to write this expansion, we have assumed that $\hat H$ is Weyl-ordered.
However, if the Hamiltonian had a different ordering, one would just need to take into account that
any power of basic operators $(\hat q_i^n \hat p_i^m)_{\rm order}$ can be written as a linear
combination of Weyl-ordered products $(\hat q_i^a \hat p_i^b)_{\rm Weyl}$ with $a=0,\dots, n$ and $b=0,\dots,m$
(check, for instance, relations \eqref{rel_ordering_1} and \eqref{rel_ordering_2} of App.~\ref{appendix_2}). 
This would simply lead to some additional terms in the expansion of the Hamiltonian above.

Now, according to \eqref{ehrenfest_2}, the equations of motion of the expectation values $(q_i,p_i)$ and
moments $\Delta(q_1^{a_1}p_1^{b_1}\dots q_N^{a_N}p_N^{b_N})$ can be given in terms of Poisson bracket, as in any Hamiltonian
system,
\begin{align}\label{quantum_equations_motion}
\begin{split}
&\frac{dq_i}{dt}=\{q_i,H_Q\},
\\
&\frac{dp_i}{dt}=\{p_i,H_Q\},
\\
&\frac{d}{dt}\Delta\big(q_1^{a_1}p_1^{b_1}\dots q_N^{a_N}p_N^{b_N}\big)=\big\{\Delta\big(q_1^{a_1}p_1^{b_1}\dots q_N^{a_N}p_N^{b_N}\big),H_Q\big\}.
\end{split}
\end{align}
Taking into account that $\{q_i,p_j\}=\delta_{ij}$ and that, 
as shown in App.~\ref{appendix_1}, the expectation values Poisson-commute with all the moments,
$\{q_i,\Delta(q_1^{a_1}p_1^{b_1}\dots q_N^{a_N}p_N^{b_N})\}=0=\{p_i,\Delta(q_1^{a_1}p_1^{b_1}\dots q_N^{a_N}p_N^{b_N})\}$,
the first two equations read
\begin{align}\label{quantum_equations_motion_class_variables_q}
&\frac{dq_i}{dt}=\frac{\partial H_Q}{\partial p_i},
\\
\label{quantum_equations_motion_class_variables_p}
&\frac{dp_i}{dt}=-\frac{\partial H_Q}{\partial q_i}.
\end{align}
These equations formally look as the Hamilton equations for the classical system,
but, instead of derivatives of the classical Hamiltonian $H$, derivatives of $H_Q$ appear on
the right-hand side. The moments that appear in $H_Q$ describe the quantum backreaction of the variables,
and in general, $(q_i,p_i)$ will not follow a classical trajectory in phase space.
The vanishing of all the moments explicitly defines the classical limit of the dynamics.
In particular, in order to study the evolution of $(q_i,p_i)$,
or of any other $\langle \hat A(\hat q_i, \hat p_i)\rangle$,
it is required to follow the evolution of all
moments simultaneously.

Concerning the moments, their evolution equations can be quite complicated
since, in general, they involve the Poisson bracket between two arbitrary moments,
\begin{align}\label{eq_motion_moments_q}
\!\frac{d}{dt}\Delta\big(q_1^{a_1}p_1^{b_1}\dots q_N^{a_N}p_N^{b_N}\big)&\!=\!
\sum_{J=2}^{+\infty}\!
\left(\prod_{k=1}^{N}\frac{1}{m_k!n_k!}\right)\!
\dfrac{\partial^{J}H}
{\partial q_1^{m_1}\partial p_1^{n_1}\dots\partial q_N^{m_N}p_N^{n_N}}
\left\{\Delta\big(q_1^{a_1}p_1^{b_1}\dots q_N^{a_N}p_N^{b_N}\big),\Delta(q_1^{m_1}p_1^{n_1}\dots q_N^{m_N}p_N^{n_N})\right\}.
\end{align}
From here, it is clear that a closed formula for such bracket is very useful
for any practical application of this formalism. Such formula was presented in Ref. \cite{BS06},
but it was not correct and, for the particular case of one degree of freedom, the correct
expression was given in Ref. \cite{Bojowald:2010qm}. In App.~\ref{appendix_2}, we generalize this result and
derive a closed formula
for the Poisson bracket between moments for any number of degrees of freedom.
The final simplified expression reads
\begin{align}\label{quantum_general_moments}
\begin{split}
&\left\{
\Delta\big(q_1^{a_1}p_1^{b_1}\dots q_N^{a_N}p_N^{b_N}\big),
\Delta\big(q_1^{m_1}p_1^{n_1}\dots q_N^{m_N}p_N^{n_N}\big)
\right\}
\\&\;=
\sum_{j=1}^{N}
\bigg(b_j\, m_j\,\Delta\big(q_1^{a_1}p_1^{b_1}\dots 
q_j^{a_j}p_j^{b_j-1}\dots q_N^{a_N}p_N^{b_N}\big)\,
\Delta\big(q_1^{m_1}p_1^{n_1}\dots 
q_j^{m_j-1}p_j^{n_j}\dots q_N^{m_N}p_N^{n_N}\big)
\\
&\;\qquad\quad\;\;-
a_j\,n_j\,\Delta\big(q_1^{a_1}p_1^{b_1}\dots 
q_j^{a_j-1}p_j^{b_j}\dots q_N^{a_N}p_N^{b_N}\big)
\,\Delta\big(q_1^{m_1}p_1^{n_1}\dots 
q_j^{m_j}p_j^{n_j-1}\dots q_N^{m_N}p_N^{n_N}\big)\bigg)    
\\
&\;+\sum_{L=0}^{M}(-1)^L
\left(\frac{
	\hbar}{2}
\right)^{2L}
K^{\alpha_1}_{a_1b_1m_1n_1}\dots
K^{\alpha_N}_{a_Nb_Nm_Nn_N}
\Delta\big(q_1^{a_1+m_1-\alpha_1}p_1^{b_1+n_1-\alpha_1}\dots 
q_N^{a_N+m_N-\alpha_N}p_N^{b_N+n_N-\alpha_N}\big),
\end{split}
\end{align}
where we have defined  
\begin{align}\label{K_def}
&K_{abmn}^{\alpha}:=
\sum_{k=0}^{\alpha}
(-1)^{k}
k!(\alpha-k)!
\binom{a}{\alpha-k}
\binom{b}{k}
\binom{m}{k}
\binom{n}{\alpha-k},
\end{align}
$2L+1:=\alpha_1+\dots+\alpha_N$ and, for each value of $L$, all possible combinations of the integers $\alpha_j\in[0,M_j]$ should be considered, with $M_j:=\min(a_j+m_j,b_j+n_j,a_j+b_j,m_j+n_j)$. The upper bound of the sum in $L$ is thus given by $M:=(-1+\sum_{j=1}^N M_j)/2$. As can be seen, the above expression is composed by two different sums. On the one hand, the first sum (in $j$) contains in general quadratic combinations of the moments and is independent of $\hbar$.
These terms come from the fact that these are central moments, which are defined as certain power of a difference, and thus
its expansion via the Newton binomial formula leads to this kind of quadratic combinations. On the other hand, the second sum
(in $L$) is linear in moments and contains even powers of $\hbar$. These terms appear in the process of reordering of the basic operators
and their origin relies in the noncommutativity of the basic operators. This is indeed a purely quantum property, as will be made explicit below
when constructing a similar formalism for classical ensembles.

In summary, this formalism replaces the wave function (or the density matrix), which depends on $N+1$ variables (time
and $q_i$ in the $q$-representation) and obeys the Schr\"odinger equation (or the von Neumann equation),
by an infinite set of moments \eqref{def_moments} and the expectation values $(q_i,p_i)$, which only depend on time.\footnote{This correspondence is exact at a formal level, but, due to the infinite size of the system of the moments, there might be convergence
issues for the evolution of quantities computed in terms of the two different
frameworks.
}
The evolution of these variables is given by
the system of equations  {\eqref{quantum_equations_motion_class_variables_q}--\eqref{eq_motion_moments_q}}. Except for certain particular Hamiltonians that will be discussed below, in general,
this system is infinite and the dynamics of all moments is highly coupled. Therefore,
it is usually complicated to obtain exact solutions. However, taking into account that the dimensions
of a given moment $\Delta(q_1^{a_1}p_1^{b_1}\dots q_N^{a_N}p_N^{b_N})$ are the same as the dimensions
of $\hbar^{(a_1+b_1+\dots+a_N+b_N)/2}$, it is usual to introduce a truncation of the system by neglecting all moments $\Delta(q_1^{a_1}p_1^{b_1}\dots q_N^{a_N}p_N^{b_N})$ such that $a_1+b_1+\dots+a_N+b_N$ is larger that a given cut off.
This is supposed to provide an accurate description of the evolution for semiclassical states
peaked on a classical trajectory in phase space.
Due to these considerations, we will refer to the sum $a_1+b_1+\dots+a_N+b_N$ as the {\it order} of the corresponding moment
$\Delta(q_1^{a_1}p_1^{b_1}\dots q_N^{a_N}p_N^{b_N})$. In this way, zeroth and first-order moments are vanishing by definition,
second-order moments are the fluctuations, $\Delta(q_i^2)$ and $\Delta(p_i^2)$, and correlations, such as $\Delta(q_i p_j)$
or $\Delta(q_iq_j)$ and $\Delta(p_ip_j)$ for $i\neq j$, while higher-order moments describe other properties of the
probability distribution under consideration.

\section{Classical ensembles}\label{sec:classical}

In Ref. \cite{BM98, Brizuela:2014cfa} it was shown that a formalism similar to the one presented above for describing quantum systems
can be constructed for classical ensembles. Here
we will provide the generalization of such formalism for an arbitrary number of degrees of freedom.

Let us thus assume a classical ensemble composed by $N$ degrees of freedom
$(\tilde q_i,\tilde p_i)_{i=1}^N$ with canonical symplectic
structure $\{\tilde q_i,\tilde p_j\}_c=\delta_{ij}$.
(Note that we are using tilde to indicate classical variables and
the subscript $c$ is used to distinguish this bracket from
the one defined previously for expectation values.)
At a certain time $t$, the probability distribution $\rho(t,\tilde q_1,\tilde p_1,\dots,\tilde q_N,\tilde p_N) d\tilde q_1d\tilde p_1\dots  d\tilde q_N d\tilde p_N$ encodes
the probability that the system is in an infinitesimal volume around the
position $(\tilde q_1,\tilde p_1,\dots,\tilde q_N,\tilde p_N)$ in phase space.
Given the Hamiltonian $H=H(\tilde q_1,\tilde p_1\dots,\tilde q_N,\tilde p_N)$, the evolution of the distribution is governed by the Liouville equation,
\begin{align}\label{Liouville}
\frac{d\rho}{dt}=
\frac{\partial\rho}{\partial t}+\{\rho,H\}_c
=0,
\end{align}
which simply states the conservation of the probability along physical trajectories.

For any function $f=f(\tilde q_1,\tilde p_1\dots,\tilde q_N,\tilde p_N)$, its classical expectation value is given by
\begin{align}\label{expectation_value_clas}
\langle f\rangle_c:=\int d\tilde q_1d\tilde p_1\dots  d\tilde q_N d\tilde p_N f(\tilde q_1,\tilde p_1,\dots,\tilde q_N,\tilde p_N)\rho(t,\tilde q_1,\tilde p_1,\dots,\tilde q_N,\tilde p_N),
\end{align}
where the integral runs over the whole phase space.
This expectation value can be used to define the central moments of the classical distribution,
\begin{align}\label{def_moments_clas}
\Delta_c\big(q_1^{a_1}p_1^{b_1}\dots q_N^{a_N}p_N^{b_N}\big):=\left\langle
(\tilde{q}_1-q_1)^{a_1}
(\tilde{p}_1-p_1)^{b_1}
\dots
(\tilde{q}_N-q_N)^{a_N}
(\tilde{p}_N-p_N)^{b_N}
\right\rangle_{c},
\end{align}
where $q_i:=\langle \tilde q_i\rangle_c$ and $p_i:=\langle \tilde p_i\rangle_c$
are the coordinates of the centroid of the distribution on the phase space.
Contrary to the quantum case, all these variables commute, and thus the ordering
inside the above expectation value is irrelevant.

In order to obtain the evolution of these classical moments,
by deriving the expression \eqref{expectation_value_clas} and using the Liouville equation \eqref{Liouville},
one can show that the time derivative of any expectation value $\langle f \rangle_c$ reads
\begin{align}\label{properties_expectation_value_clas}
\begin{split}
\frac{d\langle f\rangle_c}{dt}
&=-\int  d\tilde q_1d\tilde p_1\dots  d\tilde q_N d\tilde p_N f(\tilde q_1,\tilde p_1,\dots,\tilde q_N,\tilde p_N)\big\{\rho(t,\tilde q_1,\tilde p_1,\dots,\tilde q_N,\tilde p_N),H(t,\tilde q_1,\tilde p_1,\dots,\tilde q_N,\tilde p_N)\big\}_c.
\end{split}
\end{align}
Writing the classical Poisson bracket as derivatives with respect to the
fundamental variables $(q_i,p_i)_{i=1}^N$ and integrating by parts, this can be rewritten as
\begin{align}
\frac{d\langle f\rangle_c}{dt}&=\langle \{f,H\}_c \rangle_c.
\end{align}
Taking this result into account, it is then straightforward to define a Poisson bracket between classical expectation values,
\begin{align}\label{poisson_class_expectation_values}
\left\{\langle f\rangle_c,\langle g\rangle_c\right\}:=\langle\{f,g\}_c\rangle_c,
\end{align}
so that the dynamics is governed by the effective Hamiltonian $H_C:=\langle H\rangle_c$. That is,
the time evolution of any expectation value is given by $\frac{d}{dt}\langle f\rangle_c=\{\langle f\rangle_c,\langle H\rangle_c\}$.

As in the quantum case, we expand the effective Hamiltonian
around the centroid of the distribution, namely,
\begin{align}\label{effective_Hamiltonian_clas}
\begin{split}
H_C:=&\big\langle{H(t,\tilde  q_1,\tilde  p_1,\dots,\tilde  q_N,\tilde  p_N)}\big\rangle_{c}
\nonumber
\\
=&H(t,q_1,p_1,\dots,q_N,p_N)+\sum_{J=2}^{+\infty}
\left(\prod_{k=1}^{N}\frac{1}{m_k!}\frac{1}{n_k!}\right)
\dfrac{\partial^{J}H}
{\partial q_1^{m_1}\partial p_1^{n_1}\dots\partial q_N^{m_N}p_N^{n_N}}
\Delta_c\big(q_1^{m_1}p_1^{n_1}\dots q_N^{m_N}p_N^{n_N}\big).
\end{split}
\end{align}
The equations of motion of the expectation values and moments
are then given as follows,
\begin{align}
\begin{split}
&\frac{dq_i}{dt}:=\{q_i,H_C\}=\frac{\partial H_C}{\partial p_i},
\\
&\frac{dp_i}{dt}:=\{p_i,H_C\}=-\frac{\partial H_C}{\partial q_i},
\\
&\frac{d}{dt}\Delta_c\big(q_1^{a_1}p_1^{b_1}\dots q_N^{a_N}p_N^{b_N}\big)=\big\{\Delta_c\big(q_1^{a_1}p_1^{b_1}\dots q_N^{a_N}p_N^{b_N}\big),H_C\big\}.
\end{split}
\end{align}
In the first two equations we have taken into account that $\{q_i,p_j\}=\delta_{ij}$ and $\{\Delta_c(q_1^{a_1}p_1^{b_1}\dots q_N^{a_N}p_N^{b_N}),q_i\}=0=\{\Delta_c(q_1^{a_1}p_1^{b_1}\dots q_N^{a_N}p_N^{b_N}),p_i\}$, as in the quantum case.
These two equations show now the departure of the centroid of the distribution from the classical trajectories
on the phase space due to the distributional character of the classical ensemble. If the probability distribution
is a Dirac delta, all classical moments would be vanishing, and one would recover the usual Hamilton equations.

Following similar steps as in the quantum case (see App.~\ref{appendix_2}), the following closed formula for the Poisson bracket
between any two classical moments can be derived:
\begin{align}\label{general_formula_moments_classic}
\begin{split}
&\left\{
\Delta_c\big(q_1^{a_1}p_1^{b_1}\dots q_N^{a_N}p_N^{b_N}\big),
\Delta_c\big(q_1^{m_1}p_1^{n_1}\dots q_N^{m_N}p_N^{n_N}\big)
\right\}
\\
&\;=
\sum_{j=1}^{N}\bigg(
b_jm_j\Delta_c\big(q_1^{a_1}p_1^{b_1}\dots 
q_j^{a_j}p_j^{b_j-1}\dots q_N^{a_N}p_N^{b_N}\big)
\Delta_c\big(q_1^{m_1}p_1^{n_1}\dots 
q_j^{m_j-1}p_j^{n_j}\dots q_N^{m_N}p_N^{n_N}\big)
\\
&\;\qquad\quad\;\;-
a_jn_j\Delta_c\big(q_1^{a_1}p_1^{b_1}\dots 
q_j^{a_j-1}p_j^{b_j}\dots q_N^{a_N}p_N^{b_N}\big)
\Delta_c\big(q_1^{m_1}p_1^{n_1}\dots 
q_j^{m_j}p_j^{n_j-1}\dots q_N^{m_N}p_N^{n_N}\big)
\\
&\;\qquad\quad\;\;+
(a_jn_j-b_jm_j)\Delta_c\big(q_1^{a_1+m_1}p_1^{b_1+n_1}\dots 
q_j^{a_j+m_j-1}p_j^{b_j+n_j-1}\dots q_N^{a_N+m_N}p_N^{b_N+n_N}\big)\bigg).
\end{split}
\end{align}
It is interesting to note that
this formula can be formally obtained from the Poisson bracket between quantum moments \eqref{quantum_general_moments}
by simply imposing $\hbar=0$. More precisely, the first two terms in the sum above are the same quadratic combination
of moments that appear in \eqref{quantum_general_moments}, while the last term, linear in moments, corresponds to the
$L=0$ term of the second sum in \eqref{quantum_general_moments}.
Therefore, the equations that describe the dynamics of a quantum state and a classical ensemble
exactly coincide in the limit $\hbar=0$.
In this sense, the dynamics of a classical ensemble
can thus be understood as given by a truncated version of the quantum equations of motion.

\section{Hybrid Systems}\label{sec:hybrid}

In this section we will describe a hybrid system with mixed classical and quantum degrees of freedom in terms of its corresponding moments, as done in the previous sections for purely quantum
or classical systems.
Let us thus consider a system with $N=N_c+N_q$ degrees of freedom. The $N_c$
classical degrees of freedom will be denoted with a tilde $(\tilde{q}_k,\tilde{p}_k)_{k=1}^{N_c}$ 
and are canonically conjugate with symplectic structure $\{\tilde q_j,\tilde p_k\}_c=\delta_{jk}$.
In turn,
the quantum sector of the system is described by the basic operators $(\hat{q}_j,\hat{p}_j)_{j=N_c+1}^{N}$,
which are denoted with a hat and
follow the canonical commutation relation $[\hat{q}_j,\hat{p}_k]=i\hbar\delta_{jk}$.

In principle, the set of classical observables ${\cal A}_c$ is given by
all real functions $\tilde f=\tilde f(\tilde{q}_k,\tilde{p}_k)$ on the classical phase space,
while Hermitian (symmetric) operators
$\hat A=\hat A(\hat{q}_j,\hat{p}_j)$ on the Hilbert space form
the set of quantum observables ${\cal A}_q$.
However, since the present formalism requires an expansion in power series
of the different objects,
we need to restrict ourselves to {\it analytic} observables. That is,
${\cal A}_c$ will be formed by real analytic functions $\tilde f$ on the phase space,
and ${\cal A}_q$ will be given by Hermitian operators with an analytic dependence
on the basic operators $(\hat{q}_j,\hat{p}_j)$.
In this way, the set of hybrid observables
is defined as the direct product $\mathcal{A}:=\mathcal{A}_c\otimes\mathcal{A}_q$,
which can then be linearly spanned by powers of the basic variables of the form
$\tilde{q}_k^{a_k}\tilde{p}_k^{b_k}(\hat{q}_j^{n_j}\hat{p}_j^{m_j})_{\rm Weyl}$,
where the subscript Weyl stands here also for completely symmetric ordering.
That is, any hybrid observable $\overline{A}\in\mathcal{A}$,
which will be denoted by an overline, can be written as a linear combination,
\begin{align}\label{generic_observable}
	\overline{A}=
	\sum_{a_k b_k n_j m_j}
	C_{a_k b_k n_j m_j}\,
	\tilde{q}_1^{a_1}\tilde{p}_1^{b_1}\dots \tilde{q}_{N_c}^{a_{N_c}}\tilde{p}_{N_c}^{b_{N_c}}
	\big(\hat{q}_{N_c+1}^{n_1}\hat{p}_{N_c+1}^{m_1}\dots \hat{q}_{N}^{n_{N}}\hat{p}_{N}^{m_{N}}\big)_{\rm Weyl}	,
\end{align}
with real coefficients $C_{a_k b_k n_j m_j}$. Note that, the inclusion of non-Hermitian operators
and complex functions would be straightforward by simply allowing these coefficients to be complex.

Now, given a Hamiltonian $\overline H=\overline H(t,\tilde{q}_1,\tilde{p}_1,\dots
\tilde{q}_{N_c},\tilde{p}_{N_c},\hat{q}_{N_c+1},\hat{p}_{N_c+1},\dots
\hat{q}_{N},\hat{p}_{N})\in {\cal A}$,
we are interested in describing the evolution it generates on the expectation value
of any hybrid observable $\langle\overline A\rangle$.\footnote{We will not provide the explicit definition of the hybrid expectation value
since we will not require it. In fact, this is one of the advantages of the present formalism, since
it can be applied to different fundamental definitions.
}
Due to the decomposition \eqref{generic_observable}, it is clear that $\langle\overline A\rangle$
will be given as a linear combination of hybrid moments, which
can be formally defined in the same way as the quantum \eqref{def_moments} or classical \eqref{def_moments_clas} ones. Therefore, in order to compute the evolution of $\langle\overline A\rangle$, one just
needs to obtain the evolution of the hybrid moments.
Furthermore, following the discussion of the previous sections, the goal would to define an effective
Hamiltonian $H_{\text{eff}}:=\langle \overline H\rangle$, expand it as a power series,
\begin{align}\label{effective_Hamiltonian_hybrid}
	\begin{split}
		H_{\text{eff}}:=&\big\langle{\overline H(t,\tilde  q_1,\tilde  p_1,\dots,\tilde q_{N_c},\tilde p_{N_c},\hat q_{N_c+1},\hat  p_{N_c+1},\dots,\hat q_{N},\hat  p_{N})}\big\rangle
		\\
	=&H(t,q_1,p_1,\dots,q_N,p_N)+\sum_{J=2}^{+\infty}
	\left(\prod_{k=1}^{N}\frac{1}{m_k!}\frac{1}{n_k!}\right)
	\dfrac{\partial^{J}H}
	{\partial q_1^{m_1}\partial p_1^{n_1}\dots\partial q_N^{m_N}p_N^{n_N}}
	\Delta\big( q_1^{m_1}p_1^{n_1}\dots q_N^{m_N}p_N^{n_N}\big),
	\end{split}
\end{align}
and compute the time-derivative of any moment {as its Poisson bracket with $H_{\rm eff}$}.
Hence, the problem reduces to define a Poisson bracket between hybrid moments.

For definiteness, let us provide here the explicit expression for the hybrid moments,
 \begin{align}\label{def_hybrid_moments}
 	\Delta\big( q_1^{m_1}p_1^{n_1}\dots q_N^{m_N}p_N^{n_N}\big):=\bigg\langle
 	\prod_{i=1}^{N_c}(\tilde q_i-q_i)^{m_i}(\tilde p_i-p_i)^{n_i}
 	\prod_{j=N_c+1}^{N}(\hat q_j-q_{j})^{m_{j}}(\hat p_j-p_{j})^{n_{j}}
 	\bigg\rangle_{\text{Weyl}},
 \end{align}
where $q_i$ and $p_i$ denote the expectation value of
the corresponding classical or quantum variable, i.e.,
$q_i:=\langle\tilde q_i\rangle$ and $p_i:=\langle\tilde p_i\rangle$ 
for $i=1,\dots,N_c$, while $q_i:=\langle \hat q_i\rangle$ and $p_i:=\langle\hat p_i\rangle$
for $i=N_c+1,\dots,N$.
Taking into account that the Poisson bracket between classical moments \eqref{general_formula_moments_classic}
can be understood as a truncation of the bracket between quantum moments \eqref{quantum_general_moments},
simply by setting $\hbar=0$, one can try to prescribe a bracket for hybrid moments.
We note that,
in the last sum of \eqref{quantum_general_moments}
the combination $\alpha_i=1$ and $\alpha_j=0$ for all $j\neq i$ provides the term with $L=0$,
which is independent of $\hbar$.
Therefore, it is natural to prescribe a hybrid bracket so that every classical
degree of freedom $(\tilde q_i,\tilde p_i)$ only contributes with $\alpha_i=1$ and $\alpha_j=0$ for all $j\neq i$
to the sum in $L$ of \eqref{quantum_general_moments}, and $\alpha_i=0$ otherwise. This consideration leads to the
following definition,

\begin{align}\label{hybrid_general_moments}
\begin{split}
	&\left\{
	\Delta\big( q_1^{a_1}p_1^{b_1}\dots q_N^{a_N}p_N^{b_N}\big),
	\Delta\big(q_1^{m_1}p_1^{n_1}\dots q_N^{m_N}p_N^{n_N}\big)
	\right\}
	\\
	&\;=
	\sum_{j=1}^{N}\bigg(
	b_jm_j\Delta\big(q_1^{a_1}p_1^{b_1}\dots 
	q_j^{a_j}p_j^{b_j-1}\dots q_N^{a_N}p_N^{b_N}\big)
	\Delta\big(q_1^{m_1}p_1^{n_1}\dots 
	q_j^{m_j-1}p_j^{n_j}\dots q_N^{m_N}p_N^{n_N}\big)
	\\
	&\;\qquad\quad\;\;-
	a_jn_j\Delta\big(q_1^{a_1}p_1^{b_1}\dots 
	q_j^{a_j-1}p_j^{b_j}\dots q_N^{a_N}p_N^{b_N}\big)
	\Delta\big(q_1^{m_1}p_1^{n_1}\dots 
	q_j^{m_j}p_j^{n_j-1}\dots q_N^{m_N}p_N^{n_N}\big)
	\\
	&\;\qquad\quad\;\; +
	(a_jn_j-b_jm_j)\Delta\big(q_1^{a_1+m_1}p_1^{b_1+n_1}\dots 
	q_j^{a_j+m_j-1}p_j^{b_j+n_j-1}\dots q_N^{a_N+m_N}p_N^{b_N+n_N}\big)\bigg)
	\\
	&\;+\sum_{L=1}^{M_q}(-1)^L
\left(\frac{
		\hbar}{2}
	\right)^{2L}
	K^{\alpha_{N_c+1}}_{a_{N_c+1}b_{N_c+1}m_{N_c+1}n_{N_c+1}}\dots
	K^{\alpha_N}_{a_Nb_Nm_Nn_N}
	\\
	&\;\quad \times\Delta\big(q_1^{a_1+m_1}p_1^{b_1+n_1}\dots q_{N_c}^{a_{N_c}+m_{N_c}}p_{N_c}^{b_{N_c}+n_{N_c}}
	q_{N_c+1}^{a_{N_c+1}+m_{N_c+1}-\alpha_{N_c+1}}p_{N_c+1}^{b_{N_c+1}+n_{N_c+1}-\alpha_{N_c+1}}\dots 
	q_N^{a_N+m_N-\alpha_N}p_N^{b_N+n_N-\alpha_N}\big),
	\end{split}
\end{align}
where we have written explicitly the term $L=0$. Thus, the last sum begins now in $L=1$
and it only runs over the quantum degrees of freedom, that is, $2L+1=\alpha_{N_c+1}+\dots+\alpha_{N}$, where $\alpha_j\in[0,M_j]$ and $M_j:=\min(a_j+m_j,b_j+n_j,a_j+b_j,m_j+n_j)$. The upper bound of this sum is then given by $M_q=(-1+\sum_{j=N_c+1}^{N}M_j)/2$.
In turn, the bracket between the expectation values $(q_i,p_i)$ will be the canonical one
$\{q_i,p_j\}=\delta_{ij}$.

This bracket obeys the standard properties that one would expect.
It automatically reduces to either \eqref{quantum_general_moments} or \eqref{general_formula_moments_classic}
when all the degrees of freedom are quantum or classical, respectively.
The bracket between a pure quantum and pure classical moment is vanishing.
It is antisymmetric and the bracket between any moment and a constant is vanishing.
By construction, it is bilinear and acts as a derivative on the product between the expectation
value of two observables, that is,
\begin{equation}\label{Leibniz}
\left \{\langle\overline A\rangle, \langle\overline B\rangle\langle\overline C\rangle\right\}=
 \left\{\langle\overline A\rangle, \langle\overline B\rangle\right\}\langle\overline C\rangle+
\langle\overline B\rangle \left\{\langle\overline A\rangle, \langle\overline C\rangle\right\},
 \end{equation}
for any $\overline A, \overline B, \overline C \in {\cal A}$. And therefore,
we can define the time derivative of any $\langle\overline A\rangle$
by
\begin{equation}\label{hybrid_derivative}
 \frac{d\langle\overline A\rangle}{dt}=\big\{\langle\overline A\rangle,\langle\overline H\rangle\big\},
\end{equation}
which can be reduced to the bracket between moments \eqref{hybrid_general_moments}, by considering the expansions \eqref{generic_observable}--\eqref{effective_Hamiltonian_hybrid}
and systematically using the bilinearity and Leibniz rule \eqref{Leibniz}.
Hence, the dynamics of the expectation value of any observable and, in particular, of the hybrid
moments is set up.

Here some comments are in order though. On the one hand, strictly speaking the bracket \eqref{hybrid_general_moments} is not a Poisson bracket since
it does not obey the Jacobi identity. More precisely, 
hybrid moments, along with the expectation values $(q_i,p_i)_{i=1}^N$, form an {\it almost} Poisson algebra over the reals.
That is, the vector space over the reals spanned by the hybrid moments and expectation values
contains two bilinear products: the usual product and the bracket \eqref{hybrid_general_moments}. This bracket
obeys all the properties of a Lie bracket (antisymmetry and bilinearity), except
the Jacobi identity. However, the failure to obey the Jacobi identity is a necessary property
of any consistent bracket for hybrid systems \cite{CS99}. Quite generically, if one assumes that
there exists a bracket that obeys the Jacobi identity for a hybrid system, a number
of basic inconsistencies appear (see Ref. \cite{CS99} for some examples).
It is interesting to note that the bracket between quantum moments \eqref{quantum_general_moments}
does obey the Jacobi identity, and it is thus a Lie bracket. Furthermore, it turns out that the classical
bracket \eqref{general_formula_moments_classic} also follows the Jacobi identity, and, remarkably, it
seems to be the only truncation of \eqref{quantum_general_moments} with such property. On the other hand,
it is important to remark that this bracket satisfies the so-called \textit{definite benchmark}
proposed in Ref. \cite{Peres_Terno}, which requires that the classical limit of the
hybrid classical-quantum system and of the fully quantum system coincide.
In particular, such limit is obtained by simply imposing $\hbar= 0$ in the equations
\eqref{quantum_general_moments} and \eqref{hybrid_general_moments}, which leads to the
corresponding classical ensemble with brackets \eqref{general_formula_moments_classic}.

Note that we have derived \eqref{hybrid_general_moments} as an {\it ad hoc} truncation of
the quantum version \eqref{quantum_general_moments}.
However, in the previous sections we have used certain {\it internal} Lie bracket
$\{\cdot{},\cdot{}\}_{\rm int}$ between observables, which in the classical case is $\{\cdot{},\cdot{}\}_c$
and in the quantum case $-i\hbar^{-1}[\cdot{},\cdot{}]$,
to define the Poisson bracket between expectation values in a more fundamental way
as $\{\langle\cdot{}\rangle,\langle\cdot{}\rangle\}:=\langle \{\cdot{},\cdot{}\}_{\rm int} \rangle$.
It turns out that,
in the hybrid case, it is also possible to define the almost Poisson bracket \eqref{hybrid_general_moments}
as $\{\langle\cdot{}\rangle,\langle\cdot{}\rangle\}:=
\langle [\![\cdot{},\cdot{}]\!] \rangle$ if one prescribes the following bracket between hybrid observables,
\begin{align}\label{mixed_bracket}
\begin{split}
		&[\![\tilde f\hat{F},\tilde g\hat{G}]\!]
	:=\big\{\tilde f,\tilde g\big\}_c\,R(\hat F,\hat G)
	+\frac{1}{i\hbar}\left[\hat{F},\hat{G}\right]\tilde f\tilde g,
\end{split}
\end{align}
for any $\tilde f,\tilde g\in\mathcal{A}_c$ and
$\hat F,\hat G\in\mathcal{A}_q$, and where $R$ is a symmetric bilinear
mapping $R:\mathcal{A}_q\times\mathcal{A}_q\to\mathcal{A}_q$
with action 
\begin{equation}\label{def_R_f}
R\left(
\big(\hat q_1^{m_1}\hat p_1^{n_1}\dots
\hat q_N^{m_N}\hat p_N^{n_N}\big)_{\text{Weyl}},
\big(\hat q_1^{k_1}\hat p_1^{l_1}\dots
\hat q_N^{k_N}\hat p_N^{l_N}\big)_{\text{Weyl}}
\right):=\big(\hat q_1^{m_1+k_1}\hat p_1^{n_1+l_1}\dots
\hat q_N^{m_N+k_N}\hat p_N^{n_N+l_N}\big)_{\text{Weyl}}.
\end{equation}
Note that $R$ is defined only for Weyl-ordered products of basic quantum operators. However,
given that they have an analytic dependence on basic operators,
its action on any $\hat F$ and $\hat G$ can be computed
by performing an expansion of the form \eqref{generic_observable}.
The hybrid bracket \eqref{mixed_bracket} generalizes the quantum commutator
and the classical Poisson bracket.
As expected, in general (unless it is reduced to the trivial cases
with both observables being either purely quantum or classical),
it does not follow the Jacobi identity, and it thus defines an almost Lie algebra. In App.~\ref{app:Lie}
we derive this bracket from first principles and detail its properties.
In order to obtain \eqref{hybrid_general_moments} from \eqref{mixed_bracket}
one needs to follow the same steps as explained in App.~\ref{appendix_2} for the quantum case.

\section{Dynamics of harmonic Hamiltonians}\label{sec:harmonic}

Harmonic Hamiltonians $H$, defined as those that are at most quadratic on basic variables,
show very special dynamical properties. Since, in this case the third-order derivatives
of $H$ are vanishing, the corresponding effective Hamiltonian will be given
by a finite number of terms and will only contain moments up to second order,\footnote{Recall that we have defined
the order of a moment $\Delta(q_1^{a_1}p_1^{b_1}\dots q_N^{a_N}p_N^{b_N})$ as the sum
$a_1+b_1+\dots+a_N+b_N$.} 
\begin{equation}
H_{\rm eff}= H+ \sum_{i=1}^{N}\sum_{j=1}^{N}\frac{1}{2}\left(\frac{\partial^2 H}{\partial q_i \partial q_j}\Delta\big(q_iq_j\big)+2
\frac{\partial^2 H}{\partial q_i \partial p_j}\Delta\big(q_ip_j\big)
+\frac{\partial^2 H}{\partial p_i \partial p_j}\Delta\big(p_ip_j\big)\right),
\end{equation}
where $N$ stands for the number of degrees of freedom.
This expression is valid irrespectively of the nature (classical or quantum) of the different degrees of freedom. 

Now, it turns out that the quantum Poisson bracket \eqref{hybrid_general_moments} between a moment of order $n$ and a second-order
moment is given by a linear combination of moments of order $n$ and, furthermore, it does not contain any term
with $\hbar$. That is, if one of the entries in the bracket \eqref{hybrid_general_moments} is a second-order moment,
the sum in $L$ is reduced to $L=0$. Therefore, such bracket will have the exact same form for
moments associated to any kind (classical or quantum) of degrees of freedom.

This key property has important consequences on the dynamics described by this kind of harmonic Hamiltonians.
On the one hand, all orders decouple and the equation of motion of a given moment is coupled only to
moments of the same order.
In particular, the equations of motion for the expectation values $(q_i,p_i)$ are the classical Hamilton equations.
Thus, $(q_i,p_i)$ follow classical trajectories in phase space
and there is no backreaction produced by the moments.
On the other hand,
the system of evolution equations for the moments is linear and, since there is no $\hbar$ in any of the equations, it
has the same exact form irrespectively of the nature (classical or quantum) of the different degrees of freedom.
Therefore, given the same initial data, a classical, a quantum, or, in general, a hybrid ensemble,
will follow exactly the same dynamics and one would not be able to tell whether a given variable
is classical or quantum. At most, the character of each degree of freedom can be fixed by providing
initial data for the classical sector that are in principle not allowed for the quantum sector due to the uncertainty
relation, like
for instance being described by a Dirac delta distribution (which implies all moments to be vanishing).
Nonetheless, if the classical and quantum degrees of freedom are coupled, the dynamics will mix their
particular properties, transferring the quantumness or classicality from one sector to the other.
In particular, this will allow the {\it quantum} variables to violate the Heisenberg uncertainty relation.
As an explicit example that shows the rise of such hybrid properties, in the next section
we will study a simple hybrid harmonic Hamiltonian that describes the evolution of a classical
and a quantum coupled harmonic oscillators.

\section{Application: coupled classical and quantum harmonic oscillators
}\label{sec:application}

Let us consider two coupled one-dimensional harmonic oscillators, one classical and the other quantum,
as described by the following Hamiltonian,
\begin{align}\label{Hamiltonian_oscillator}
	\overline H=\frac{1}{2}(\tilde p^2+\omega^2\tilde q^2)
	+\frac{1}{2}(\hat k^2+\omega^2\hat x^2)+\gamma \tilde q\hat x,
\end{align}
where $\omega> 0$ is the frequency and $\gamma> 0$ the coupling constant, which, without loss of generality,
are taken to be positive.
Following the notation of the previous section, $(\tilde q,\tilde p)$ and $(\hat x, \hat k)$
correspond to the canonically conjugate position and momentum of the classical and quantum sectors, respectively.
This system was already studied in Ref.~\cite{BCG12} in terms of Wigner distributions.
Here we will make a more explicit and systematic study of the evolution of the moments, though the main qualitative conclusions will be similar as those obtained there.
Note that, in order to perform a detailed analytic analysis,
we are also restricting ourselves to the case with both
oscillators having the same frequency. This is, as explained in Ref.~\cite{BCG12}, the situation
where the transfer of quantum uncertainty between oscillators turns out to be most efficient
and hybrid properties of the system surface in a clear way.

\begin{figure}[t]
	\centering
	\begin{subfigure}[h]{0.46\textwidth}
	\centering
	\includegraphics[width=\linewidth]{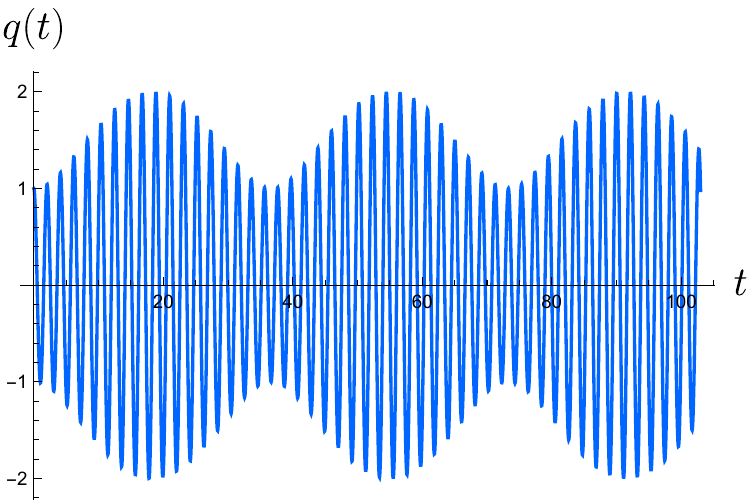}
	\end{subfigure}
	\hfill
	\begin{subfigure}[h]{0.46\textwidth}
		\centering
		\includegraphics[width=\textwidth]{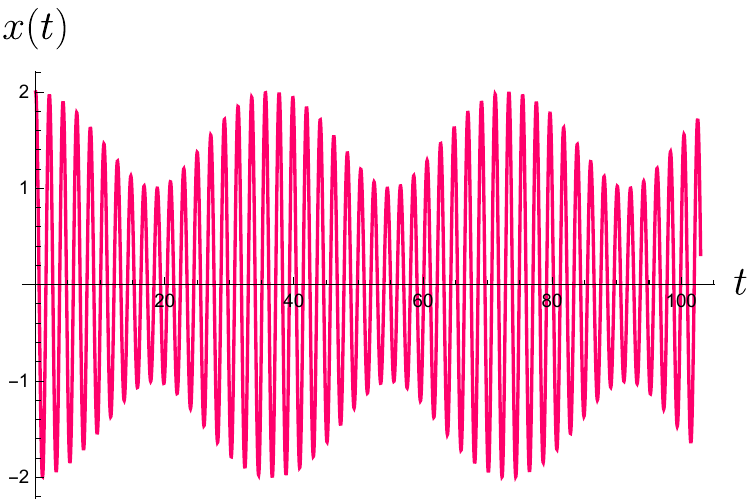}
	\end{subfigure}
	\caption{Evolution of the positions $q$ and $x$ of each oscillator in the weak-coupling regime for $\omega_1=3$, $\omega_2=2\sqrt{2}$ ($\omega=\sqrt{17/2}$ and $\gamma=1/2$), and initial conditions $q_0=1$, $x_0=2$, and $p_0=k_0=0$.}
	\label{fig:evolution_beat}
\end{figure}

The dynamics of the system is then governed by the effective
Hamiltonian $\langle \overline H\rangle$, which, expanded in moments, reads as
\begin{align}\label{eff_Hamiltonian_oscillator}
	H_{H}=\langle \overline H\rangle =\frac{1}{2}( p^2+\omega^2 q^2)
	+\frac{1}{2}(k^2+\omega^2 x^2)+\gamma qx+
	\frac{1}{2}\left( \Delta(p^2)+\omega^2 \Delta(q^2)\right)
	+\frac{1}{2}\left(\Delta(k^2)+\omega^2 \Delta(x^2)\right)+\gamma \Delta( q x),
\end{align}
where $q:=\langle\tilde q\rangle$, $p:=\langle\tilde p\rangle$, $x:=\langle\hat  x\rangle$, and $k:=\langle\hat  k\rangle$.
Since the Hamiltonian \eqref{Hamiltonian_oscillator} is quadratic in basic variables,
as explained in the previous section, equations at different orders decouple and, in particular, expectation values
of basic variables obey the classical equations of motion,
\begin{align}
\frac{dq}{dt} &= p, \\
\frac{dp}{dt}&= -\omega q-\gamma x,\\
\frac{dx}{dt}&= k,\\
\frac{dk}{dt}&= -\omega x-\gamma p.
\end{align}
The solution to these equations can be written as follows,
\begin{align}\label{sol_q_oscillator}
q=&\frac{1}{2}\left[
(q_0+x_0)\cos\left(\omega_1t\right)+
(q_0-x_0)\cos\left(\omega_2t\right)
+\frac{k_0+p_0}{\omega_1}\sin\left(\omega_1t\right)
+\frac{p_0-k_0}{\omega_2}\sin\left(\omega_2t\right)
\right],
\\\label{sol_p_oscillator}
p=&\frac{1}{2}\left[
(p_0+k_0)\cos\left(\omega_1t\right)+
(p_0-k_0)\cos\left(\omega_2t\right)
-(q_0+x_0)\omega_1\sin\left(\omega_1t\right)
+(x_0-q_0)\omega_2\sin\left(\omega_2t\right)
\right],
\\
\label{sol_x_oscillator}
x=&\frac{1}{2}\left[
(q_0+x_0)\cos\left(\omega_1t\right)+
(x_0-q_0)\cos\left(\omega_2t\right)
+\frac{k_0+p_0}{\omega_1}\sin\left(\omega_1t\right)
+\frac{k_0-p_0}{\omega_2}\sin\left(\omega_2t\right)
\right],
\\
\label{sol_k_oscillator}
k=&\frac{1}{2}\left[
(p_0+k_0)\cos\left(\omega_1t\right)+
(k_0-p_0)\cos\left(\omega_2t\right)
-(q_0+x_0)\omega_1\sin\left(\omega_1t\right)
+(q_0-x_0)\omega_2\sin\left(\omega_2t\right)
\right],
\end{align}
where $q_0$, $p_0$, $x_0$, and $k_0$ are the initial values of the different variables at $t=0$,
and we have defined $\omega_1:=\sqrt{\omega^2+\gamma}$ and $\omega_2:=\sqrt{\omega^2-\gamma}$. Since this evolution is identical to the one followed by two classical
(with a pointlike, Dirac-delta, distribution in phase space)
coupled oscillators, a priori it does not present any specific particularity due to hybridization.
On the one hand,
in the weak-coupling regime $(\gamma<\omega^2)$, both $\omega_1$ and $\omega_2$ are real,
and the dynamics is oscillatory. In the particular case 
$x_0=q_0$ and $k_0=p_0$, both oscillators oscillate symmetrically with a frequency $\omega_1$,
while for $x_0=-q_0$ and $k_0=-p_0$, the oscillations are antisymmetric and have frequency $\omega_2$.
However, excluding these particular cases, in general the evolution is periodic
only when $\omega_1$ and $\omega_2$ are commensurable, i.e., $\omega_2/\omega_1=m/n\in\mathbb{Q}$.
In such a case, the period of the oscillations is $2\pi n/\omega_1=2\pi n/\sqrt{\omega^2+\gamma}$.
Additionally, it is interesting to note that, in the weak-coupling regime ($\gamma\ll\omega^2$), the frequencies become very similar, $\omega_1\approx\omega_2$. This leads to the occurrence of beats, where the evolutions are approximately sinusoidal with a slowly varying amplitude, as depicted in Fig.~\ref{fig:evolution_beat}. On the other hand,
in the strong-coupling regime $(\gamma>\omega^2)$ $\omega_2$ is purely imaginary and
an exponentially growing mode appears, which is only suppressed for the trajectory
with $x_0=q_0$ and $k_0=p_0$, where both oscillators oscillate with frequency $\omega_1$ as one and the same.
The transition between the two regimes (with $\gamma=\omega^2$) leads to a dynamics with an
oscillating mode of frequency $\sqrt{2}\omega$ and a linearly increasing or decreasing mode.
In what follows, let us assume $\gamma<\omega^2$ and thus consider the oscillatory regime,
since it presents more interesting properties.

The equations of motion for the moments can be obtained by computing the Poisson bracket
with the effective Hamiltonian \eqref{eff_Hamiltonian_oscillator}, namely,
\begin{align}\label{oscillator_eqm_moments}
\begin{split}
\frac{d}{dt}\Delta\big(q^{m_1}p^{n_1}x^{m_2}k^{n_2}\big)&=m_1\Delta\big(q^{m_1-1}p^{n_1+1}x^{m_2}k^{n_2}\big)
-\omega^2n_1\Delta\big(q^{m_1+1}p^{n_1-1}x^{m_2}k^{n_2}\big)
+m_2\Delta\big(q^{m_1}p^{n_1}x^{m_2-1}k^{n_2+1}\big)
\\
&-\omega^2n_2\Delta\big(q^{m_1}p^{n_1}x^{m_2+1}k^{n_2-1}\big)
-\gamma n_1\Delta\big(q^{m_1}p^{n_1-1}x^{m_2+1}k^{n_2}\big)
-\gamma n_2\Delta\big(q^{m_1+1}p^{n_1}x^{m_2}k^{n_2-1}\big).
\end{split}
\end{align}
As commented in the previous section for quadratic Hamiltonians, these equations are linear
and only couple moments of the same order. Moreover,
it turns out that it is possible to write the solution of the full system
in the following compact form:
\begin{align}\label{sol_moments_oscillator}
\Delta\big(q^{m_1}p^{n_1}x^{m_2}k^{n_2}\big)=&
\sum_{j_1+j_2+j_3+j_4={S}}^S\,
\frac{C_{j_1 j_2 j_3 j_4}}{j_1!j_2!j_3!j_4!}
\dfrac{\partial^{S}(q^{m_1}p^{n_1}x^{m_2}k^{n_2})}{\partial q_0^{j_1}\partial p_0^{j_2}\partial x_0^{j_3}\partial k_0^{j_4}},
\end{align}
where $S=m_1+n_1+m_2+n_2$ is the order of the moment under consideration and
$C_{j_1j_2j_3j_4}:=\Delta(\tilde q^{j_1}\tilde p^{j_2}\hat x^{j_3}\hat k^{j_4})|_{t=0}$ encode the initial data at $t=0$.
Note also that, in the argument of the derivative $q^{m_1}p^{n_1}x^{m_2}k^{n_2}$, one should replace
the corresponding solution \eqref{sol_q_oscillator}--\eqref{sol_k_oscillator}. In this way,
since the evolution of the moments is
expressed as a sum of partial derivatives of the expectation values $q$, $x$, $p$ and $k$,
these are also periodic when $\omega_1$ and $\omega_2$ are commensurable.

As mentioned above, since the Hamiltonian is quadratic,
the equations of motion for the different variables are unaffected by the nature of the different degrees of freedom.
That is, it makes no difference whether they are quantum or classical, and thus
equations \eqref{eff_Hamiltonian_oscillator}--\eqref{sol_moments_oscillator} above describe the evolution of two coupled harmonic oscillators
(which can be both classical, both quantum, or one classical and one quantum).
Nevertheless, the range of possible values for the initial moments $C_{m_1n_1m_2n_2}$ varies in each scenario due to the requirement imposed by the Heisenberg uncertainty principle
on the quantum variables.
In the case with two classical oscillators, there is not such a restriction and all the moments can be vanishing,
but with two quantum oscillators or in the hybrid scenario, Heisenberg uncertainty relation does not allow for an exact vanishing
of all moments.

\begin{figure}[t]
	\centering
	\includegraphics[width=0.8\linewidth]{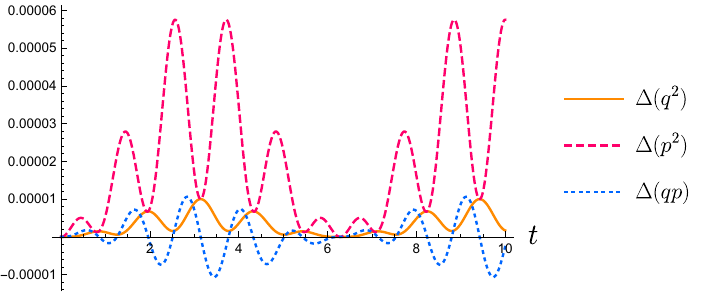}
	\caption{Evolution of the second-order moments of the classical sectors when they are initially vanishing, for $\omega_1=3$, $\omega_2=2$ ($\omega=\sqrt{13/2}$ and $\gamma=5/2$), $C_{1100}=0$, $C_{2000}=10^{-5}$ and $C_{0200}=10^{-5}$.}
	\label{fig:oscillator_initially_vanishing}
\end{figure}

Therefore, this hybrid system can exhibit unique properties not present in classical-classical or quantum-quantum coupled oscillators.
For instance, even if we choose initially vanishing fluctuations for the classical variables,
$C_{m_1n_1m_2n_2}=0$ for all cases with $m_1+n_1\neq 0$, classical moments are later activated due to quantum fluctuations.
The dynamics for such scenario is described by simply imposing the commented initial data in the above expression,
\begin{align}\label{sol_moments_ini_oscillator}
\Delta\big(q^{m_1}p^{n_1}x^{m_2}k^{n_2}\big)=&
\sum_{j_1+j_2={S}}^S\,
\frac{C_{00j_1 j_2}}{j_1!j_2!}
\dfrac{\partial^{S}(q^{m_1}p^{n_1}x^{m_2}k^{n_2})}{\partial  x_0^{j_1}\partial k_0^{j_2}},
\end{align}
From here it is possible to check that none of the moments are exactly vanishing
for $t>0$, and even pure classical moments $\Delta(q^{m_1} p^{n_1})$
are activated. In order to see this fact more clearly, let us, for instance,
write explicitly the evolution of the second-order moments for the classical sector:
\begin{align}
\nonumber
\Delta(q^2)&=\frac{1}{4}\left[
\cos\left(\omega_1 t\right)-
\cos\left(\omega_2 t
\right)\right]^2C_{0020}
+\frac{1}{4}\left[
\frac{\sin\left(\omega_1 t\right)}{\omega_1}-\frac{\sin\left(\omega_2 t\right)}{\omega_2}
\right]^2
C_{0002}
\\
&+
\frac{1}{2}
\left[
\cos\left(\omega_1 t\right)-
\cos\left(\omega_2 t
\right)\right]
\left[
\frac{\sin\left(\omega_1 t\right)}{\omega_1}-\frac{\sin\left(\omega_2 t\right)}{\omega_2}
\right]
C_{0011},
\label{class_fluctuations_ini_q2_oscillator}
\\
\nonumber
\Delta(p^2)&=
\frac{1}{4}
\left[
\omega_2\sin\left(\omega_2 t\right)-\omega_1\sin\left(\omega_1 t\right)
\right]^2
C_{0020}
+\frac{1}{4}\left[
\cos\left(\omega_1 t\right)-
\cos\left(\omega_2 t
\right)\right]^2C_{0002}
\\
&+\frac{1}{2}
\left[
\cos\left(\omega_1 t\right)-
\cos\left(\omega_2 t
\right)\right]
\left[
\omega_2\sin\left(\omega_2 t\right)-\omega_1\sin\left(\omega_1 t\right)
\right]
C_{0011},
\label{class_fluctuations_ini_p2_oscillator}
\\
\nonumber
\Delta(qp)&=
\frac{1}{4}
\left[
\omega_2\sin\left(\omega_2 t\right)-\omega_1\sin\left(\omega_1 t\right)
\right]\left[
\cos\left(\omega_1 t
\right)-\cos\left(\omega_2 t\right)\right]
C_{0020}
\\&
+\frac{1}{4}\left[
\cos\left(\omega_1 t\right)-
\cos\left(\omega_2 t
\right)\right]
\left[
\frac{\sin\left(\omega_1 t\right)}{\omega_1}-\frac{\sin\left(\omega_2 t\right)}{\omega_2}
\right]
C_{0002}
\label{class_fluctuations_ini_qp_oscillator}
\\
&+\frac{1}{4}
\left[
\cos\left(2\omega_1 t\right)+\cos\left(2\omega_2 t\right)-2\cos\left(\omega_1 t\right)\cos\left(\omega_2 t\right)
+\frac{(\omega_1^2+\omega_2^2)}{\omega_1\omega_2}
\sin\left(\omega_1 t\right)\sin\left(\omega_2 t\right)
\right]C_{0011}.
\nonumber
\end{align}
From these expressions it is clear that they are all initially vanishing but activate for $t>0$.
For illustrational purposes, their evolution is also shown in Fig.~\ref{fig:oscillator_initially_vanishing}. 

Moreover, another important property relies on the exchange of quantumness and classicality
between different sectors. In order to analyze how the expression of the
Heisenberg uncertainty relation evolves, we define
the uncertainty corresponding to each degree of freedom as follows,
\begin{align}
U_q:=&\Delta(x^2)\Delta(k^2)-\Delta(xk)^2,\\
U_c:=&\Delta(q^2)\Delta(p^2)-\Delta(qp)^2.
\end{align}
For a purely quantum system $U_q\geq \hbar^2/4$ would be obeyed all along evolution, while
for a purely classical system one would simply observe $U_c\geq 0$. However, in this hybrid case,
considering again, for simplicity, that all the moments of the classical sector are initially vanishing,
from \eqref{sol_moments_oscillator}, one gets
\begin{align}\label{Uncertainty_clas}
	&U_c=\frac{U_0}{16\omega_1^2\omega_2^2}\left(
	2\omega_1\omega_2-g(t)
	\right)^2,
	\\
	\label{Uncertainty_quan}
	&U_q=\frac{U_0}{16\omega_1^2\omega_2^2}\left(
	2\omega_1\omega_2+g(t)
	\right)^2,
\end{align}
with $U_0:=C_{0020}C_{0002}-C_{0011}^2\geq \hbar^2/4$ being the initial uncertainty of the quantum
sector, and $g$ a function defined as

\begin{align}\label{g}
	g(t):=2\omega_1\omega_2\cos(\omega_1 t)\cos(\omega_2 t)+\left(\omega_1^2+\omega_2^2\right)\sin(\omega_1 t)\sin(\omega_2 t).
\end{align}
Therefore, dynamical properties of $U_c$ and $U_q$ rely on those of $g(t)$:
both $U_c$ and $U_q$ oscillate, and, if $g(t)$ is periodic, they are also periodic, which, once again, corresponds to the case with $\omega_2/\omega_1\in\mathbb{Q}$. In addition, the upper bounds of the uncertainties are finite and are determined by the singular points of $g(t)$, 
\begin{align}\label{max_uncertainty}
	\begin{split}
	&U_c\leq \frac{U_0}{16\omega_1^2\omega_2^2}\left(
	2\omega_1\omega_2-g_{\min}
	\right)^2=:U_c^{\max},
	\\
	&U_q\leq \frac{U_0}{16\omega_1^2\omega_2^2}\left(
	2\omega_1\omega_2+g_{\max}
	\right)^2=:U_q^{\max},
	\end{split}
\end{align}
where $g_{\min}$ and $g_{\max}$ correspond to the minimum and maximum values of $g(t)$, respectively.
Unfortunately,
it is not possible to obtain a general expression for these values for arbitrary $\omega_1$ and $\omega_2$.
However, it is possible to see that they are bounded as follows,
\begin{align}\label{max_min_g}
\begin{split}
	&g_{\min}\in\left[-\left(\omega_1^2+\omega_2^2\right),-2\omega_1\omega_2\right],\\
& g_{\max}\in\left[2\omega_1\omega_2,\omega_1^2+\omega_2^2\right],
\end{split}
\end{align} 
which provides an interval for the maximum value of the uncertainties,
\begin{align}\label{max_uncertainty_2}
	U_c^{\max},U_q^{\max}\in\left[U_0,\frac{\left(\omega_1+\omega_2\right)^4}{16\omega_1^2\omega_2^2}U_0\right].
\end{align}
Moreover,
from \eqref{max_min_g} we deduce that the interval $[-2\omega_1\omega_2,2\omega_1\omega_2]$ is in the image of $g$, and, consequently, according to \eqref{Uncertainty_clas} and \eqref{Uncertainty_quan}, the uncertainties must vanish at some point. That is, their lower-bound is zero,
\begin{align}\label{min_uncertainty}
0\leq U_c\leq U_c^{\max},\\
0\leq U_q\leq U_q^{\max},
\end{align}
and thus the quantum sector violates, in this way, the Heisenberg uncertainty relation.
In fact,
from \eqref{Uncertainty_clas}--\eqref{Uncertainty_quan} one can see that the sum
between the uncertainties reads,
\begin{equation}
U_c+U_q=\frac{U_0}{2}\left(1+\frac{g^2}{4\omega_1^2\omega_2^2} \right),
\end{equation}
which has a minimum bound at $U_0/2$ and thus it obeys
\begin{equation}\label{eq.totaluncertainty}
U_c+U_q\geq U_0/2.
\end{equation}
This inequality can be interpreted as the uncertainty relation of the full
hybrid system: even if the Heisenberg relation of the quantum sector
$U_q\geq\hbar^2/4$ is allowed to be violated by its coupling to the classical oscillator,
the sum between the uncertainties $U_c+U_q$ has a minimum positive bound,
which still encodes certain limit in the precision on the
simultaneous knowledge of different physical properties of the system.
It is interesting to note that if, instead of this hybrid system,
one considered two couple quantum oscillators initially saturating both the uncertainty relation,
i.e., $U_0=\hbar^2/4$ for each one, during evolution their combined uncertainty would have a minimum of $2 U_0$,
while the hybrid system presents a minimum of $U_0/2$.
Therefore, we can state that the classical sector has two diminishing effects
on the total uncertainty of the system: it does not contribute to it (since initially we can take $U_c=0$),
but, in addition, it allows the quantum sector to decrease its own, which leads to a one fourth
reduction as compared to the quantum-quantum system.

\begin{figure}[t]
	\centering
	\includegraphics[width=1.0\linewidth]{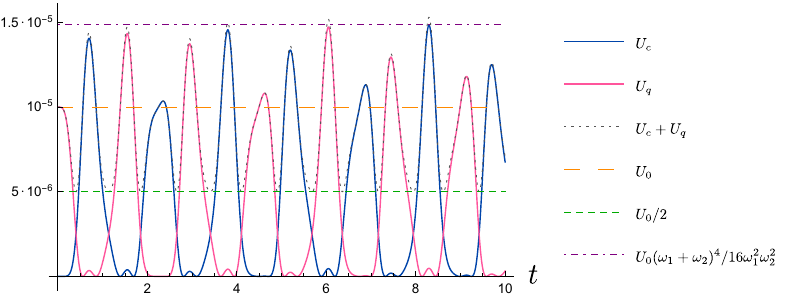}
	\caption{Evolution of the classical $U_c$ (solid in blue) and quantum $U_q$ (solid in pink) uncertainties, together with their sum $U_c+U_q$ (in gray and dotted), for $\omega_1=2\sqrt{2}$, $\omega_2=7$ ($\omega=\sqrt{57/2}$ and $\gamma=41/2$), and initial $U_0=10^{-5}$. The long dashed orange line corresponds to the initial value $U_0$, the dashed green line to the minimum value of $U_c+U_q$, given by $U_0/2$, and the dot-dashed purple one to the greatest possible upper bound for the uncertainties $U_0(\omega_1+\omega_2)^4/16\omega_1^2\omega_2^2$.}
	\label{fig:uncertainty}
\end{figure}

Finally, from \eqref{Uncertainty_clas}--\eqref{Uncertainty_quan} one can also obtain the following
interesting relation
\begin{equation}\label{eq.conservation}
(U_q-U_c- U_0)^2-4\, U_0\, U_c=0,
\end{equation}
which is obeyed all along evolution. We interpret this as the conservation
of the total quantumness (or classicality) of the full system.
In particular, from here it is straightforward to see that, when
the uncertainty of one of the sectors ($U_c$ or $U_q$)
vanishes, the uncertainty of the other sector takes the value $U_0$, though this is not necessarily its maximum.
The evolution of the uncertainty of each sector, $U_q$ and $U_c$, and of their sum $U_c+U_q$ is depicted in Fig. \ref{fig:uncertainty}
for certain particular values of the parameters and initial conditions.

\section{Conclusions}\label{sec:conclusions}

We have presented a complete and consistent formalism to describe the dynamics of
quantum systems, classical ensembles and, in general, hybrid systems with mixed classical
and quantum sectors, for an arbitrary number of degrees of freedom.
This framework is based on the decomposition
of the probability distribution function of the system into its infinite set of moments.
For quantum systems this formalism was already presented in Ref.~\cite{BS06}, though here
we have derived the closed formula \eqref{quantum_general_moments} for the Poisson bracket between any two
quantum moments, which corrects the expression given in that reference.
(The correct expression for the case with one degree of freedom was already presented in Ref. \cite{Bojowald:2010qm}.)

In addition, generalizing the work of Refs.~\cite{BM98, Brizuela:2014cfa} to any number of degrees of freedom,
we have shown that in this setup the dynamics of a classical ensemble can be described in a very similar
way as that of a quantum system.
In fact, the Poisson bracket between classical moments \eqref{general_formula_moments_classic}
takes the same form as its quantum version with $\hbar=0$.
This restriction essentially drops all the purely quantum terms that come from the noncommutativity of the basic operators.
In this sense, the dynamics of a classical ensemble can thus be understood as given
by a truncated version of the quantum equations of motion.

In the hybrid case observables are given by products between
Hermitian operators on the Hilbert space and real functions on the phase space. For such systems,
starting from first principles, we have
defined the hybrid bracket between observables \eqref{mixed_bracket}, which generalizes the quantum commutator
and the classical Poisson bracket. It turns out that this hybrid bracket obeys all the properties
(bilinearity and antisymmetry)
of a Lie bracket, except the Jacobi identity,
and therefore it defines an almost Lie algebra. Then, making use of such bracket
between observables, one can define a bracket for the expectation values of hybrid observables that,
for the moments, takes the explicit form \eqref{hybrid_general_moments}. As one would expect, this bracket does
not follow the Jacobi identity either, and, along with the usual product, it provides an almost Poisson algebra.
Remarkably, this bracket between hybrid moments can be derived
from the bracket between quantum moments
\eqref{quantum_general_moments} simply by neglecting the $\hbar$ terms that would be present if
a given classical degree of freedom was quantum.

In summary, this framework can be used to study the evolution of any hybrid system
described in terms of conjugate variables and a Hamiltonian.
As compared to other approaches, one of the advantages is that,
unlike the probability distribution,
the moments, being expectation values of observables, are physical quantities accessible by experiments.
Furthermore, in this setup, it is not necessary to explicitly
construct the mathematical structures on the Hilbert space and the phase space
(or any other space one considers to describe the dynamics of the hybrid system),
since one directly works with expectation values.

As an application of the formalism, in the last section we have studied in detail a particular
hybrid system given by two (a quantum and a classical) harmonic oscillators. These oscillators
are coupled by a quadratic interaction term, which allows for classical
and quantum properties to be transferred between them. In particular, we have shown that the quantum oscillator
is allowed to violate the Heisenberg uncertainty relation, as long as the uncertainty
in the classical oscillator is not vanishing. In fact, as suggested in Ref. \cite{BCG12}, one can define
a combined uncertainty of the oscillators, which has a minimum positive bound. We have obtained
this bound explicitly for the present model, and one can interpret the inequality \eqref{eq.totaluncertainty}
as the uncertainty relation of the full hybrid system.
Interestingly, this analysis shows that the coupling of the quantum oscillator to the classical
system has two diminishing effects in the total uncertainty of the system. On the one hand,
the classical sector does not contribute to the total uncertainty. But, on the other hand,
it also allows the quantum sector to decrease its own uncertainty. This leads to a reduction
of a fourth on the total uncertainty of the hybrid system, as compared to the system
given by two quantum oscillators with an initial minimum uncertainty. Finally, we have also obtained the relation
\eqref{eq.conservation} that can be understood as the conservation law of the total
quantumness or classicality of the system.

\section*{Acknowledgments}

This work is supported by
the Basque Government Grant \mbox{IT1628-22}, and
by the Grant PID2021-123226NB-I00 (funded by MCIN/AEI/10.13039/501100011033 and by “ERDF A way of making Europe”).
SFU is funded by an FPU fellowship of the Spanish Ministry of Universities.

\appendix

\section{General properties of the Poisson bracket between expectation values}\label{app:Poisson}

In this appendix, we will prove general properties of the Poisson bracket between expectation values. More precisely, in Sec.~\ref{appendix_1} we will show that the Poisson bracket between
the expectation value of the basic operators and any moment is vanishing.
In Sec.~\ref{appendix_2} we will derive the general formula \eqref{quantum_general_moments} for the Poisson bracket between moments.

\subsection{Poisson bracket between expectation values of the basic operators and moments}\label{appendix_1}

Here our goal is to show that the bracket 
$\{q_i,\Delta(q_1^{a_1}p_1^{b_1}\dots q_N^{a_N}p_N^{b_N})\}$ and
$\{p_i,\Delta(q_1^{a_1}p_1^{b_1}\dots q_N^{a_N}p_N^{b_N})\}$ are vanishing.
We will present the explicit computations for the former, since for the latter
the derivation is equivalent.

First, we expand the expression of the moment by making use of Newton's binomial
formula, and use the linearity in the Poisson bracket to write,
\begin{align}\label{app_poisson_exp_moments_1}
\begin{split}
	\big\{q_i,\Delta\big(q_1^{a_1}p_1^{b_1}\dots q_N^{a_N}p_N^{b_N}\big)\big\}=&\sum_{k_1=0}^{a_1}\sum_{l_1=0}^{b_1}\dots\sum_{k_N=0}^{a_N}\sum_{l_N=0}^{b_N}
\left(
\prod_{j=1}^N
\binom{a_j}{k_j}\binom{b_j}{l_j}
(-1)^{k_j+l_j}
\right)
\\
&\times\big\{
q_i,q_1^{k_1}p_1^{l_1}\dots q_N^{k_N}p_N^{l_N}\langle\hat q_1^{a_1-k_1}
\hat p_1^{b_1-l_1}
\dots\hat q_N^{a_N-k_N}
\hat p_N^{b_N-l_N}
\rangle_{\text{Weyl}} \big\}.
\end{split}
\end{align}
Then, by applying the Leibniz rule and taking into account that $\{q_i,q_j\}=0$ and $\{q_i,p_j\}=\delta_{ij}$,
\begin{align}\label{app_poisson_exp_moments_2}
\begin{split}
&\big\{
q_i,q_1^{k_1} p_1^{l_1}\dots q_N^{k_N} p_N^{l_N}\langle\hat q_1^{a_1-k_1}
\hat p_1^{b_1-l_1}
\dots\hat q_N^{a_N-k_N}
\hat p_N^{b_N-l_N}
\rangle_{\text{Weyl}}\big\}
\\
&\,=l_iq_i^{k_i}p_i^{l_i-1}
\langle\hat q_1^{a_1-k_1}
\hat p_1^{b_1-l_1}
\dots\hat q_N^{a_N-k_N}
\hat p_N^{b_N-l_N}
\rangle_{\text{Weyl}}
\prod_{\substack{j=1\\j\neq i}}^{N}
q_j^{k_j}p_j^{l_j}+
\big\{
q_i,\langle\hat q_1^{a_1-k_1}
\hat p_1^{b_1-l_1}
\dots\hat q_N^{a_N-k_N}
\hat p_N^{b_N-l_N}
\rangle_{\text{Weyl}}
\big\}\prod_{j=1}^N q_j^{k_j}p_j^{l_j}.
\end{split}
\end{align}
Using now the definition of the Poisson bracket between expectation values, $\{\langle \hat f\rangle,\langle \hat g\rangle\}:=-i\hbar^{-1}\langle[\hat f,\hat g]\rangle$, and considering that different degrees of freedom commute,
\begin{align}\label{app_poisson_exp_moments_3}
\begin{split}
\big\{
q_i,\langle\hat q_1^{a_1-k_1}
\hat p_1^{b_1-l_1}
\dots\hat q_N^{a_N-k_N}
\hat p_N^{b_N-l_N}
\rangle_{\text{Weyl}}
\big\}:=&\frac{1}{i\hbar}\bigg\langle
\left[\hat q_i,\big(\hat q_1^{a_1-k_1}
\hat p_1^{b_1-l_1}
\dots\hat q_N^{a_N-k_N}
\hat p_N^{b_N-l_N}
\big)_{\text{Weyl}}\right]
\bigg\rangle
\\
=&\frac{1}{i\hbar}\bigg\langle
\prod_{\substack{j=1\\j\neq i}}^N
\big(\hat{q}_j^{a_j-k_j}\hat{p}_j^{b_j-l_j}\big)_{\text{Weyl}}
\left[
\hat q_i,\big(\hat{q}_i^{a_i-k_i}\hat{p}_i^{b_i-l_i}\big)_{\text{Weyl}}
\right]
\bigg\rangle
\\
=&(b_i-l_i)\langle
\hat{q}_1^{a_1-k_1}\hat{p}_1^{b_1-l_1}
\dots\hat{q}_i^{a_i-k_i}\hat{p}_i^{b_i-l_i-1}\dots \hat{q}_N^{a_N-k_N}\hat{p}_N^{b_N-l_N}
\rangle_{\text{Weyl}},
\end{split}
\end{align}
where we have taken into account that 
\begin{align*}
	\left[
	\hat q_i,\big(\hat{q}_i^{a_i-k_i}\hat{p}_i^{b_i-l_i}\big)_{\text{Weyl}}
	\right]=
	i\hbar(b_i-l_i)\big(\hat{q}_i^{a_i-k_i}\hat{p}_i^{b_i-l_i-1}\big)_{\text{Weyl}}.
\end{align*}
Then, combining the results \eqref{app_poisson_exp_moments_1}--\eqref{app_poisson_exp_moments_3},
the bracket is given as
\begin{align*}
	\begin{split}
	\big\{q_i,\Delta\big(q_1^{a_1}p_1^{b_1}\dots q_N^{a_N}p_N^{b_N}\big)\big\}&=\sum_{k_1=0}^{a_1}\sum_{l_1=0}^{b_1}\dots\sum_{k_N=0}^{a_N}\sum_{l_N=0}^{b_N}
	\left(
	\prod_{j=1}^N
	\binom{a_j}{k_j}\binom{b_j}{l_j}
	(-1)^{k_j+l_j}
	\right)
	\\
	&\quad\times
	l_iq_1^{k_1}p_1^{l_1}\dots q_i^{k_i}p_i^{l_i-1}\dots q_N^{k_N}p_1^{l_N}
	\langle\hat q_1^{a_1-k_1}
	\hat p_1^{b_1-l_1}
	\dots\hat q_N^{a_N-k_N}
	\hat p_N^{b_N-l_N}
	\rangle_{\text{Weyl}}
	\\
	&+\sum_{k_1=0}^{a_1}\sum_{l_1=0}^{b_1}\dots\sum_{k_N=0}^{a_N}\sum_{l_N=0}^{b_N}
	\left(
	\prod_{j=1}^N
	\binom{a_j}{k_j}\binom{b_j}{l_j}
	(-1)^{k_j+l_j}
	\right)
	\\
	&\quad\times
(b_i-l_i)q_1^{k_1}p_1^{l_1}\dots q_N^{k_N}p_N^{l_N}\langle
\hat{q}_1^{a_1-k_1}\hat{p}_1^{b_1-l_1}
\dots\hat{q}_i^{a_i-k_i}\hat{p}_i^{b_i-l_i-1}\dots \hat{q}_N^{a_N-k_N}\hat{p}_N^{b_N-l_N}
\rangle_{\text{Weyl}}.
	\end{split}
\end{align*}

Finally, simplifying the binomial coefficients and performing a change of variables $l_i'=l_i-1$ in the first term, it is straightforward to obtain
\begin{align}\label{app_poisson_exp_moments_final_q}
\begin{split}
\big\{q_i,\Delta\big(q_1^{a_1}p_1^{b_1}\dots q_N^{a_N}p_N^{b_N}\big)\big\}=0.
\end{split}
\end{align}
Same rationale can be applied to prove that 
\begin{align}\label{app_poisson_exp_moments_final_p}
\begin{split}
\big\{p_i,\Delta\big(q_1^{a_1}p_1^{b_1}\dots q_N^{a_N}p_N^{b_N}\big)\big\}=0.
\end{split}
\end{align}

\subsection{Appendix: Poisson bracket between moments}\label{appendix_2}

In this appendix, we will explain the key steps and properties to obtain the general formula
\eqref{quantum_general_moments} for the Poisson bracket between two arbitrary moments
$
\{\Delta(q_1^{a_1}p_1^{b_1}\dots q_N^{a_N}p_N^{b_N}),\Delta(q_1^{m_1}p_1^{n_1}\dots q_N^{m_N}p_N^{n_N})\}.$

Let us first present some relations that will be useful along the computation.
On the one hand, from relation \eqref{app_poisson_exp_moments_3}, and following the same reasoning,
it is possible to obtain,
\begin{align}\label{app_poisson_general_4}
\begin{split}
\big\{
p_i,\langle\hat q_1^{a_1-k_1}
\hat p_1^{b_1-l_1}
\dots\hat q_N^{a_N-k_N}
\hat p_N^{b_N-l_N}
\rangle_{\text{Weyl}}
\big\}
=&-(a_i-k_i)\langle
\hat{q}_1^{a_1-k_1}\hat{p}_1^{b_1-l_1}
\dots\hat{q}_i^{a_i-k_i-1}\hat{p}_i^{b_i-l_i}\dots \hat{q}_N^{a_N-k_N}\hat{p}_N^{b_N-l_N}
\rangle_{\text{Weyl}}.
\end{split}
\end{align}
On the other hand, making use of the following properties for reordering quantum
operators,
\begin{align}
\big(\hat{q}_j^m\hat{p}_j^n\big)_{\text{Weyl}}&=
\sum_{k=0}^{\min({m,n})}
\left(\frac{i\hbar}{2}\right)^k
k!
\binom{n}{k}
\binom{m}{k}
\hat{p}_j^{n-k}
\hat{q}_j^{m-k}
\nonumber\\[10pt]
&=
\sum_{k=0}^{\min({m,n})}(-1)^k
\left(\frac{i\hbar}{2}\right)^k
k!
\binom{n}{k}
\binom{m}{k}
\hat{q}_j^{m-k}
\hat{p}_j^{n-k},
\end{align}
\begin{align}\label{rel_ordering_1}
&\hat{p}_j^n\hat{q}_j^m=\sum_{k=0}^{\min({m,n})}(-1)^k
\left(\frac{i\hbar}{2}\right)^k
k!\binom{n}{k}\binom{m}{k}
\big(\hat{q}_j^{m-k}\hat{p}_j^{n-k}\big)_{\text{Weyl}},
\\[10pt]
\label{rel_ordering_2}
&\hat{q}_j^m\hat{p}_j^n=\sum_{k=0}^{\min({m,n})}
\left(\frac{i\hbar}{2}\right)^k
k!\binom{n}{k}\binom{m}{k}
\big(\hat{q}_j^{m-k}\hat{p}_j^{n-k}\big)_{\text{Weyl}},
\end{align}
\begin{align}
\big[\hat{q}_j^m,
\hat{p}_j^n\big]=
\sum_{k=1}^{\min({m,n})}
\left(i\hbar\right)^k
k!\binom{n}{k}\binom{m}{k}
\hat{p}_j^{n-k}\hat{q}_j^{m-k}
=
-\sum_{k=1}^{\min({m,n})}
(-1)^k\left(i\hbar\right)^k
k!\binom{n}{k}\binom{m}{k}
\hat{q}_j^{m-k}\hat{p}_j^{n-k},
\end{align}
it is possible to obtain the key relations,

\begin{align}\label{app_aid1}
\big(\hat{q}_j^m\hat{p}_j^n
\big)_{\text{Weyl}}
\big(\hat{q}_j^s\hat{p}_j^r
\big)_{\text{Weyl}}
&=\sum_{\alpha=0}^{M'}
\bigg(
\frac{i
	\hbar}{2}
\bigg)^{\alpha}
K^{\alpha}_{mnsr}
\big(\hat{q}_j^
{m+s-\alpha}\hat{p}_j^{n+r-\alpha}
\big)_{\text{Weyl}}
,
\\
\label{app_aid2}
\left[\big(\hat{q}_j^m\hat{p}_j^n
\big)_{\text{Weyl}},
\big(\hat{q}_j^s\hat{p}_j^r
\big)_{\text{Weyl}}\right]
&=\sum_{\text{odd\;}\alpha=1}^{M'}
2
\bigg(
\frac{i
	\hbar}{2}
\bigg)^{\alpha}
K^{\alpha}_{mnsr}
\big(\hat{q}_j^
{m+s-\alpha}\hat{p}_j^{n+r-\alpha}
\big)_{\text{Weyl}}
,
\end{align}
with $M':=\min(m+s,m+n,n+r,s+r)$ and
\begin{align}\label{app_K_def}
&K_{abmn}^{\alpha}:=
\sum_{k=0}^{\alpha}
(-1)^{k}
k!(\alpha-k)!
\binom{a}{\alpha-k}
\binom{b}{k}
\binom{m}{k}
\binom{n}{\alpha-k}.
\end{align}

Now we are in a position to compute the bracket $
\{\Delta(q_1^{a_1}p_1^{b_1}\dots q_N^{a_N}p_N^{b_N}),\Delta(q_1^{m_1}p_1^{n_1}\dots q_N^{m_N}p_N^{n_N})\}.$ First, we expand one of the moments by applying Newton's binomial formula, and
using linearity of the Poisson bracket, we write
\begin{align}\label{app_poisson_general_1}
\begin{split}
		&\left\{\Delta\big(q_1^{a_1}p_1^{b_1}\dots q_N^{a_N}p_N^{b_N}\big),\Delta\big(q_1^{m_1}p_1^{n_1}\dots q_N^{m_N}p_N^{n_N}\big)\right\}
		\\&\;=
		\sum_{r_1=0}^{m_1}\sum_{s_1=0}^{n_1}\dots\sum_{r_N=0}^{m_N}\sum_{s_N=0}^{n_N}
		\left(
		\prod_{j=1}^N
		\binom{m_j}{r_j}\binom{n_j}{s_j}
		(-1)^{r_j+s_j}
		\right)
		\\&\;\,\quad\times
	\big\{\Delta\big(q_1^{a_1}p_1^{b_1}\dots q_N^{a_N}p_N^{b_N}\big)
		,
		q_1^{r_1}p_1^{s_1}\dots q_N^{r_N}p_N^{s_N}
		\langle\hat q_1^{m_1-r_1}
		\hat p_1^{n_1-s_1}
		\dots\hat q_N^{m_N-r_N}
		\hat p_N^{n_N-s_N}
		\rangle_{\text{Weyl}} 
		\big\}.
\end{split}
\end{align}
Then, by applying the Leibniz rule and taking into account that expectation values $q_i$ and $p_i$ have a vanishing Poisson bracket with any moment (relations \eqref{app_poisson_exp_moments_final_q} and \eqref{app_poisson_exp_moments_final_p}), 
\begin{align}\label{app_poisson_general_2}
\begin{split}
\left\{\Delta\big(q_1^{a_1}p_1^{b_1}\dots q_N^{a_N}p_N^{b_N}\big),\Delta\big(q_1^{m_1}p_1^{n_1}\dots q_N^{m_N}p_N^{n_N}\big)\right\}
=&
\sum_{r_1=0}^{m_1}\sum_{s_1=0}^{n_1}\dots\sum_{r_N=0}^{m_N}\sum_{s_N=0}^{n_N}
\left(
\prod_{j=1}^N
\binom{m_j}{r_j}\binom{n_j}{s_j}
(-1)^{r_j+s_j}q_j^{r_j}p_j^{s_j}
\right)
\\&\times
\big\{\Delta\big(q_1^{a_1}p_1^{b_1}\dots q_N^{a_N}p_N^{b_N}\big)
,
\langle\hat q_1^{m_1-r_1}
\hat p_1^{n_1-s_1}
\dots\hat q_N^{m_N-r_N}
\hat p_N^{n_N-s_N}
\rangle_{\text{Weyl}} 
\big\}.
\end{split}
\end{align}
Now, we expand the other moment by using Newton's binomial formula, and considering
the linearity of the bracket, we get
\begin{align}\label{app_poisson_general_3}
\begin{split}
		&\big\{\Delta\big(q_1^{a_1}p_1^{b_1}\dots q_N^{a_N}p_N^{b_N}\big)
	,
	\langle\hat q_1^{m_1-r_1}
	\hat p_1^{n_1-s_1}
	\dots\hat q_N^{m_N-r_N}
	\hat p_N^{n_N-s_N}
	\rangle_{\text{Weyl}} 
	\big\}
	\\&\,=
	\sum_{k_1=0}^{a_1}\sum_{l_1=0}^{b_1}\dots
	\sum_{k_N=0}^{a_N}\sum_{l_N=0}^{b_N}
	\left(
	\prod_{j=1}^N
	\binom{a_j}{k_j}\binom{b_j}{r_j}
	(-1)^{k_j+r_j}
	\right)
	\\&\,\quad\times
	\big\{q_1^{k_1}p_1^{l_1}\dots q_N^{k_N}p_N^{l_N}
	\langle\hat q_1^{a_1-k_1}
	\hat p_1^{b_1-r_1}
	\dots\hat q_N^{a_N-k_N}
	\hat p_N^{b_N-r_N}
	\rangle_{\text{Weyl}} 
	,
	\langle\hat q_1^{m_1-r_1}
	\hat p_1^{n_1-s_1}
	\dots\hat q_N^{m_N-r_N}
	\hat p_N^{n_N-s_N}
	\rangle_{\text{Weyl}} 
	\big\}.
\end{split}
\end{align}
Next, by applying the Leibniz rule, and taking into account
\eqref{app_poisson_exp_moments_3} and \eqref{app_poisson_general_4},
\begin{align}
\label{app_poisson_general_5}
\begin{split}
&\big\{q_1^{k_1}p_1^{l_1}\dots q_N^{k_N}p_N^{l_N}
\langle\hat q_1^{a_1-k_1}
\hat p_1^{b_1-r_1}
\dots\hat q_N^{a_N-k_N}
\hat p_N^{b_N-r_N}
\rangle_{\text{Weyl}} 
,
\langle\hat q_1^{m_1-r_1}
\hat p_1^{n_1-s_1}
\dots\hat q_N^{m_N-r_N}
\hat p_N^{n_N-s_N}
\rangle_{\text{Weyl}} 
\big\}
\\
&\,=
\sum_{j=1}^N
\langle\hat q_1^{a_1-k_1}
\hat p_1^{b_1-r_1}
\dots\hat q_N^{a_N-k_N}
\hat p_N^{b_N-r_N}
\rangle_{\text{Weyl}}
\prod_{\substack{i=1\\i\neq j}}^N q_i^{k_i}p_i^{l_i}
\\&
\,\quad\times
\bigg(k_j(n_j-s_j)q_j^{k_j-1}p_j^{l_j}\langle\hat q_1^{m_1-r_1}
\hat p_1^{n_1-s_1}
\dots\hat q_j^{m_j-r_j}
\hat p_j^{n_j-s_j-1}
\dots\hat q_N^{m_N-r_N}
\hat p_N^{n_N-s_N}
\rangle_{\text{Weyl}} 
\\
&\,\qquad\quad-
l_j(m_j-r_j)q_j^{k_j}p_j^{l_j-1}\langle\hat q_1^{m_1-r_1}
\hat p_1^{n_1-s_1}
\dots\hat q_j^{m_j-r_j-1}
\hat p_j^{n_j-s_j}
\dots\hat q_N^{m_N-r_N}
\hat p_N^{n_N-s_N}
\rangle_{\text{Weyl}} 
\bigg)
\\
&\,+q_1^{k_1}p_1^{l_1}\dots q_N^{k_N}p_N^{l_N}\big\{
\langle\hat q_1^{a_1-k_1}
\hat p_1^{b_1-r_1}
\dots\hat q_N^{a_N-k_N}
\hat p_N^{b_N-r_N}
\rangle_{\text{Weyl}} 
,
\langle\hat q_1^{m_1-r_1}
\hat p_1^{n_1-s_1}
\dots\hat q_N^{m_N-r_N}
\hat p_N^{n_N-s_N}
\rangle_{\text{Weyl}} 
\big\}.
	\end{split}
\end{align}
Then, by applying the definition of the Poisson bracket between expectation values, $\{\langle\hat f\rangle,\langle \hat{g}\rangle\}:=-i\hbar^{-1}\langle[\hat f,\hat g]\rangle$, taking into account that different degrees of freedom commute and the
properties \eqref{app_aid1}--\eqref{app_aid2}, the last line of \eqref{app_poisson_general_5}
can be written as
\begin{align}\label{app_poisson_general_6}
\begin{split}
	&\big\{
	\langle\hat q_1^{a_1-k_1}
	\hat p_1^{b_1-r_1}
	\dots\hat q_N^{a_N-k_N}
	\hat p_N^{b_N-r_N}
	\rangle_{\text{Weyl}} 
	,
	\langle\hat q_1^{m_1-r_1}
	\hat p_1^{n_1-s_1}
	\dots\hat q_N^{m_N-r_N}
	\hat p_N^{n_N-s_N}
	\rangle_{\text{Weyl}} 
	\big\}
	\\
	&\,=\sum_{L=0}^{M''}(-1)^L
	\left(\frac{
		\hbar}{2}
	\right)^{2L}
	\prod_{j=1}^N
	K^{\alpha_j}_{a_j-k_j,b_j-r-j,m_j-r_j,n_j-s_j}
	\\
	&\,\quad\times
	\langle\hat q_1^{a_1-k_1+m_1-r_1-\alpha_1}
	\hat p_1^{b_1-l_1+n_1-s_1-\alpha_1}
	\dots\hat q_N^{a_N-k_N+m_N-r_N-\alpha_N}
	\hat p_N^{b_N-l_N+n_N-s_N-\alpha_N}
	\rangle_{\text{Weyl}} ,
	\end{split}
\end{align}
where we have also considered the Leibniz rule inside the commutator. Here, we define 
$2L+1:=\alpha_1+\dots+\alpha_N$ and, for each value of $L$, we must consider all possible combinations of the integers $\alpha_j\in[0,M'_j]$, with $M'_j:=\min(a_j+m_j-k_j-r_j,b_j+n_j-l_j-s_j,a_j+b_j-k_j-l_j,m_j+n_j-r_j-s_j)$. 

Finally, by combining \eqref{app_poisson_general_1}--\eqref{app_poisson_general_6}, and doing some combinatorics
with the series, one gets the result
\begin{align}\label{app_quantum_general_moments}
\begin{split}
&\left\{
\Delta\big(q_1^{a_1}p_1^{b_1}\dots q_N^{a_N}p_N^{b_N}\big),
\Delta\big(q_1^{m_1}p_1^{n_1}\dots q_N^{m_N}p_N^{n_N}\big)
\right\}
\\&\;=
\sum_{j=1}^{N}
\bigg(b_j\, m_j\,\Delta\big(q_1^{a_1}p_1^{b_1}\dots 
q_j^{a_j}p_j^{b_j-1}\dots q_N^{a_N}p_N^{b_N}\big)\,
\Delta(q_1^{m_1}p_1^{n_1}\dots 
q_j^{m_j-1}p_j^{n_j}\dots q_N^{m_N}p_N^{n_N})
\\
&\;\qquad\quad\;\;-
a_j\,n_j\,\Delta\big(q_1^{a_1}p_1^{b_1}\dots 
q_j^{a_j-1}p_j^{b_j}\dots q_N^{a_N}p_N^{b_N}\big)
\,\Delta(q_1^{m_1}p_1^{n_1}\dots 
q_j^{m_j}p_j^{n_j-1}\dots q_N^{m_N}p_N^{n_N})\bigg)    
\\
&\;+\sum_{L=0}^{M}(-1)^L
\left(\frac{
	\hbar}{2}
\right)^{2L}
K^{\alpha_1}_{a_1b_1m_1n_1}\dots
K^{\alpha_N}_{a_Nb_Nm_Nn_N}
\Delta\big(q_1^{a_1+m_1-\alpha_1}p_1^{b_1+n_1-\alpha_1}\dots 
q_N^{a_N+m_N-\alpha_N}p_N^{b_N+n_N-\alpha_N}\big),
\end{split}
\end{align}
where $2L+1:=\alpha_1+\dots+\alpha_N$ and, for each value of $L$, all possible combinations of the integers $\alpha_j\in[0,M_j]$ should be considered, with,  $M_j:=\min(a_j+m_j,b_j+n_j,a_j+b_j,m_j+n_j)$.
Thus, the upper bound of the sum in $L$ is thus given by $M:=(-1+\sum_{j=1}^N M_j)/2$.

\section{The almost Lie hybrid bracket}\label{app:Lie}

Let us provide a systematic construction of the hybrid bracket
$[\![\overline A,\overline B]\!]$ between two generic hybrid observables
$\overline A,\overline B\in {\cal A}$ beginning from first principles.
First, these being two fundamental properties of any Lie bracket,
we will require it to be bilinear and antisymmetric:
\begin{align}
\label{properties_anti} [\![\overline A,\overline B]\!]&=-[\![\overline B,\overline A]\!],\\
\label{properties_linear}
[\![a_1\overline A_1+a_2\overline A_2,\overline B]\!]&=a_1[\![\overline A_1,\overline B]\!]
+a_2[\![\overline A_2,\overline B]\!],
\end{align}
for any real numbers
$a_1,a_2$. Then, as suggested in Ref. \cite{GS17},
we will also require that
\begin{align}
	\label{properties_Hermitian}
[\![\overline A,\overline B]\!]^{\dagger}&=[\![\overline A,\overline B]\!], 
\end{align}
with $^{\dagger}$ being the complex conjugate, which 
implies that an Hermitian operator (an observable) will remain so under dynamical evolution, and
\begin{align}
	\label{properties_bracket_mix}
	[\![\tilde f,\tilde g\hat{G}]\!]=\big\{\tilde f,\tilde g\big\}_c\hat{G},&\qquad [\![\hat F,\tilde g\hat G]\!]=\frac{1}{i\hbar}\left[\hat F,\hat G\right]\tilde g,
\end{align}
for $\forall \tilde f,\tilde g\in\mathcal{A}_c$ and $\forall \hat F ,\hat G\in\mathcal{A}_q$.
This last requirement, combined with the bilinearity \eqref{properties_linear},
dictates several properties in a very compact way:
\begin{itemize}
	\item The bracket must reduce to the classical Poisson bracket or the commutator (times $-i/\hbar$) when both arguments are
	classical or quantum observables, respectively, 
	\begin{align*}
	[\![\tilde f,\tilde g]\!]=\big\{\tilde f,\tilde g\big\}_c,
	&\qquad
	[\![\hat F,\hat G]\!]=\frac{1}{i\hbar}
	\left[\hat F,\hat G\right].
	\end{align*}
	\item The bracket between a classical and a quantum operator vanishes:
	\begin{align*}
	[\![\tilde f,\hat Q]\!]=\big\{\tilde f,1\big\}_c\hat Q=0.
	\end{align*}
	Hence, when the quantum and classical sectors are dynamically uncoupled, i.e., when the Hamiltonian is of the form $\overline H=\hat{Q}+\tilde C$, the different sectors evolve independently; in particular, an observable of the form $\tilde f\hat G$ will evolve to $\tilde f(t)\hat G(t)$.
	\item Since the identity $\mathds{1}$ is simultaneously a classical and a quantum observable:
	\begin{align*}
	[\![\mathds{1},\tilde g\hat G]\!]=\big\{\mathds{1},\tilde g\big\}_c\hat G=0.
	\end{align*}
	Consequently, a constant operator will not evolve.
\end{itemize}
Therefore, taking the properties \eqref{properties_anti}--\eqref{properties_bracket_mix} into account,
we propose the following definition:
\begin{align}\label{mixed_bracket_def}
\begin{split}
		&[\![\tilde f\hat{F},\tilde g\hat{G}]\!]
	:=\big\{\tilde f,\tilde g\big\}_cR(\hat F,\hat G)
	+\frac{1}{i\hbar}\left[\hat{F},\hat{G}\right]\tilde f\tilde g,
\end{split}
\end{align}
where $\tilde f,\tilde g\in\mathcal{A}_c$ and $\hat F,\hat G\in\mathcal{A}_q$. As evident, definition \eqref{mixed_bracket_def} applies specifically to just two products; however, by \textit{imposing} bilinearity to the hybrid bracket, it can be extended to encompass any pair of hybrid observables (since any of these can be expressed as a linear combination of products $\tilde f_i\hat F_j$). $R$ is a
mapping $R:\mathcal{A}_q\times\mathcal{A}_q\to\mathcal{A}_q$ and, in what follows, we will analyze the conditions that it has to satisfy so that properties \eqref{properties_anti}--\eqref{properties_bracket_mix} are ensured. 
In what follows, $\hat F,\hat G,\hat F_1,\hat F_2\in\mathcal{A}_q$ will denote any generic operators in ${\cal A}_q$, $\tilde f,\tilde g\in\mathcal{A}_c$ any classical observables, and $a_1,a_2\in\mathbb{R}$ any real numbers.

\begin{enumerate}[label=$\roman*$/]
	\item First, since both the commutator and the classical Poisson bracket are antisymmetric,
	for the hybrid bracket \eqref{mixed_bracket_def} to be antisymmetric \eqref{properties_anti},
the mapping $R$ needs to be symmetric:
	\begin{align}\label{cond_R_sym}
		R(\hat F,\hat G)=R(\hat G,\hat F).
	\end{align}
	\item Second, since the hybrid bracket has to be linear \eqref{properties_linear} in its first argument, it must obey
	\begin{align*}
		[\![\tilde f(a_1\hat F_1+a_2\hat F_2),\tilde g\hat G]\!]=a_1[\![\tilde f\hat F_1,\tilde g\hat G]\!]+a_2[\![\tilde f\hat F_2,\tilde g\hat G]\!].
	\end{align*}
	Then, by making use of definition \eqref{mixed_bracket_def} in both sides of the last equation, one gets
	\begin{align*}
		&\big\{\tilde f,\tilde g\big\}_cR(a_1\hat F_1+a_2\hat F_2,\hat G)+a_1\frac{1}{i\hbar}\left[\hat F_1,\hat G_1\right]\tilde f\tilde g+a_2\frac{1}{i\hbar}\left[\hat F_2,\hat G_1\right]\tilde f\tilde g
		\\
		&=a_1\big\{\tilde f,\tilde g\big\}_cR(\hat F_1,\hat G)+a_1\frac{1}{i\hbar}\left[\hat F_1,\hat G_1\right]\tilde f\tilde g+a_2\big\{\tilde f,\tilde g\big\}_cR(\hat F_2,\hat G)+a_2\frac{1}{i\hbar}\left[\hat F_2,\hat G_1\right]\tilde f\tilde g
	,
	\end{align*}
	where we have used that the commutator is bilinear. Then, simplifying and grouping terms, one gets that 
	\begin{align}\label{cond_R_linear1}
		R(a_1\hat F_1+a_2\hat F_2,\hat G)=a_1R(\hat F_1,\hat G)+a_2R(\hat F_2,\hat G).
	\end{align}
	The same reasoning can be made for linearity in the second argument of the bracket and obtain
	\begin{align}\label{cond_R_linear2}
	R(\hat G,a_1\hat F_1+a_2\hat F_2)=a_1R(\hat G,\hat F_1)+a_2R(\hat G,\hat F_2).
	\end{align}
	Hence, the bracket \eqref{mixed_bracket_def} will be bilinear, and thus obey \eqref{properties_linear}, if $R$ is also bilinear.
	\item Third, due to condition \eqref{properties_Hermitian},
	$[\![\tilde f\hat{F},\tilde g\hat{G}]\!]$ has to be Hermitian, that is,
	\begin{align*}
	[\![\tilde f\hat{F},\tilde g\hat{G}]\!]=[\![\tilde f\hat{F},\tilde g\hat{G}]\!]^{\dagger}.
	\end{align*}
	Using definition \eqref{mixed_bracket_def}, and taking into account that $\frac{1}{i\hbar}[\hat{F},\hat{G}]$ is Hermitian,
	one gets that
	\begin{align}\label{cond_R_hermitic}
		R(\hat F,\hat G)^{\dagger}=R(\hat F,\hat G).
	\end{align}
	\item Fourth, since the identity $\hat{\mathds{1}}$ commutes with any operator, $[ \hat{\mathds{1}},\hat{G}]=0$,
	and using the first equality of \eqref{properties_bracket_mix},
	one can write
	\begin{align*}
	\big\{\tilde f,\tilde g\big\}_c\hat G\stackrel{\eqref{properties_bracket_mix}}{=}[\![\tilde f,\tilde g\hat{G}]\!]
	&:=\big\{\tilde f,\tilde g\big\}_cR(\hat{\mathds{1}},\hat G)
	+\frac{1}{i\hbar}\left[ \hat{\mathds{1}},\hat{G}\right]\tilde f\tilde g=\big\{\tilde f,\tilde g\big\}_cR(\hat{\mathds{1}},\hat G).
	\end{align*}
	Thus, condition \eqref{properties_bracket_mix} imposes that $\hat{\mathds{1}}$ must be the identity for the mapping $R$:
	\begin{align}\label{cond_R_identity}
		R(\hat{\mathds{1}},\hat G)=\hat G.
	\end{align}
	Finally, the second equality of condition \eqref{properties_bracket_mix}, $
	[\![\hat F,\tilde g\hat{G}]\!]
	:=-i\hbar^{-1}[ \hat{F},\hat{G}]\tilde g$, is directly satisfied by definition \eqref{mixed_bracket_def}
	since $\{\mathds{1},\tilde g\}=0$.
\end{enumerate}
In summary, the hybrid bracket \eqref{mixed_bracket_def} will obey the basic conditions \eqref{properties_anti}--\eqref{properties_bracket_mix}
if the mapping $R$ is symmetric \eqref{cond_R_sym}, bilinear \eqref{cond_R_linear1}--\eqref{cond_R_linear2}, Hermitian \eqref{cond_R_hermitic}, and the identity operator is also its identity \eqref{cond_R_identity}.

Let us thus now turn our attention to the last fundamental requirement for \eqref{mixed_bracket_def} to be a Lie bracket: the Jacobi identity. 
As explained in the text, being a hybrid bracket, it cannot obey such identity since this would produce certain inconsistencies
in the dynamics (see \cite{CS99}). However, since we still have some freedom to define $R$, one can use the failure
to fulfill the Jacobi identity as a guidance or additional requirement. More precisely, we define the Jacobiator of
the hybrid bracket as,
\begin{align}\label{Jacobi_def}
	\mathcal{J}:=
	[\![\overline A,[\![\overline B,\overline C]\!]]\!]+[\![\overline B,[\![\overline C,\overline A]\!]]\!]+[\![\overline C,[\![\overline A,\overline B]\!]]\!],
\end{align}
for certain hybrid observables $\overline A,\overline B,\overline C\in\mathcal{A}$. Now,
making use of definition \eqref{mixed_bracket_def}, taking bilinearity into account, and considering
$\overline A=\tilde a\hat A$, $\overline B=\tilde b\hat B$ and $\overline C=\tilde c\hat C$,
the Jacobiator reads as
\begin{align}\label{jacobiator}
\begin{split}
\mathcal{J}&=
R\left(\hat A,R(\hat B,\hat C)\right)\big\{\tilde a,\big\{\tilde b,\tilde c\big\}_c\big\}_c
+R\left(\hat B,R(\hat C,\hat A)\right)\big\{\tilde b,\big\{\tilde c,\tilde a\big\}_c\big\}_c+R\left(\hat C,R(\hat A,\hat B)\right)\big\{\tilde c,\big\{\tilde a,\tilde b\big\}_c\big\}_c
\\
&\quad+\frac{1}{i\hbar}
\tilde a\big\{
\tilde b,\tilde c
\big\}_c\left(
\big[\hat A,R(\hat B,\hat C)\big]+
R\big(\hat B,\big[\hat C,\hat A\big]\big)-R\big(\hat C,\big[\hat A,\hat B\big]\big)
\right)
\\
&\quad+\frac{1}{i\hbar}
\tilde b\big\{
\tilde c,\tilde a
\big\}_c\left(
\big[\hat B,R(\hat C,\hat A)\big]+
R\big(\hat C,\big[\hat A,\hat B\big]\big)-R\big(\hat A,\big[\hat B,\hat C\big]\big)
\right)
\\
&\quad+\frac{1}{i\hbar}
\tilde c\big\{
\tilde a,\tilde b
\big\}_c\left(
\big[\hat C,R(\hat A,\hat B)\big]+
R\big(\hat A,\big[\hat B,\hat C\big]\big)-R\big(\hat B,\big[\hat C,\hat A\big]\big)
\right),
\end{split}
\end{align}
where we have also taken into account that the commutator satisfies the Jacobi identity.
From here one can see that there are two types of terms in $J$.
The first three terms are products between operators and double classical Poisson brackets.
In fact, if one requests that
\begin{align}\label{cond_R_asso}
	R\left(\hat A,R(\hat B,\hat C)\right)=R\left(\hat B,R(\hat C,\hat A)\right)=R\left(\hat C,R(\hat A,\hat B)\right),
\end{align} 
and use the fact that the classical Poisson bracket obeys the Jacobi identity, these first three terms
would vanish. Therefore, \eqref{cond_R_asso} is a natural and simple condition for $R$
that we will also require.
The remaining terms in \eqref{jacobiator} are more involved since they mix the action of $R$ and the commutator,
so it is not possible to obtain a simple requirement for $R$ so that they all would be vanishing.
These are indeed the terms that are responsible for the Jacobiator not to be vanishing in the hybrid case.
Note, however, that if one computes the Jacobiator for purely quantum (with $\tilde a=\tilde b=\tilde c=1$) or purely classical
(with $\hat A=\hat B=\hat C=\hat{\mathds{1}}$) observables, the expression \eqref{jacobiator} would automatically vanish and the Jacobi identity
would be fulfilled.

Therefore, taking all the discussion above into account, we propose the following action
for the mapping $R$:
\begin{align}\label{def_R_app}
R\left(
\big(\hat q_1^{m_1}\hat p_1^{n_1}\dots
\hat q_N^{m_N}\hat p_N^{n_N}\big)_{\text{Weyl}},
\big(\hat q_1^{k_1}\hat p_1^{l_1}\dots
\hat q_N^{k_N}\hat p_N^{l_N}\big)_{\text{Weyl}}
\right):=\big(\hat q_1^{m_1+k_1}\hat p_1^{n_1+l_1}\dots
\hat q_N^{m_N+k_N}\hat p_N^{n_N+l_N}\big)_{\text{Weyl}},
\end{align}
which satisfies conditions \eqref{cond_R_sym}--\eqref{cond_R_identity} and \eqref{cond_R_asso}.
Making then use of the hybrid bracket \eqref{mixed_bracket_def} given by this definition of $R$, it is possible to show that
the Poisson bracket between expectation values, defined as $\{\langle\cdot{}\rangle,\langle\cdot{} \rangle\}:=\langle[\![\cdot{},\cdot{}]\!] \rangle$,
leads to the explicit form \eqref{hybrid_general_moments} for the case of two generic moments.Note that this definition of $R$ is given only for Weyl-ordered products of basic quantum operators.
However, as commented in the main text, since it is bilinear, its action can be extended to any arbitrary
operator by considering an expansion of the form \eqref{generic_observable}.

\end{document}